\documentclass{article}

\usepackage[page,header]{appendix}
\usepackage[dvipsnames]{xcolor}
\usepackage{natbib, amssymb, amsmath, amsthm, times, graphicx, tikz, comment}
\usepackage{thmtools}
\usepackage{algorithm}
\usepackage[noend]{algpseudocode}
\usepackage{booktabs, caption, makecell}
\usepackage{subfig}
\usepackage{authblk}
\usepackage{scrextend}
\usepackage{hyperref}
\usepackage{enumitem}
\usepackage[margin=3.5cm]{geometry}
\usepackage[parfill]{parskip}
\usepackage{rotating}
\usepackage{threeparttable}

\usetikzlibrary{datavisualization, plotmarks, arrows.meta, shadings}

\definecolor{color1}{RGB}{246, 193, 66} 
\definecolor{color2}{RGB}{117, 250, 98} 
\definecolor{color3}{RGB}{114, 247, 253} 
\definecolor{color4}{RGB}{10, 65, 245} 
\definecolor{color5}{RGB}{119, 37, 245} 
\definecolor{color6}{RGB}{234, 55, 181} 

\newcommand\rainbowbullet{\tikz{\path[draw=black, shade, shading=color wheel white center] (0,0) circle (2.5pt);}}
\newcommand\colouredbullet[1]{\tikz{\path[draw=black, fill=#1] (0,0) circle (2.5pt);}}

\newcommand{\To}{\textbf{ to }}
\newcommand{\Let}[2]{\State #1 $\gets$ #2}

\theoremstyle{plain}

\theoremstyle{definition}
\newtheorem{query}{Query}
\theoremstyle{remark}
\newtheorem*{remark}{Remark}

\providecommand{\keywords}[1]
{
  \small	
  \textbf{\textit{Keywords:}} #1
}

\title{Adaptive Cluster Thresholding with Spatial Activation Guarantees Using All-resolutions Inference}

\author[1,2]{Xu Chen \thanks{Alphabetic author order was used.}\thanks{Corresponding author. Einthovenweg 20, 2333 ZC Leiden, the Netherlands; X.Chen.MS@lumc.nl.}}
\author[1]{Jelle J. Goeman}
\author[3]{Thijmen J. P. Krebs}
\author[4]{Rosa J. Meijer}
\author[2]{Wouter D. Weeda}
\affil[1]{Department of Biomedical Data Sciences, Leiden University Medical Center, Leiden, Netherlands}
\affil[2]{Methodology and Statistics Unit, Institute of Psychology, Leiden University, Leiden, The Netherlands}
\affil[3]{Delft University of Technology, Delft, The Netherlands}
\affil[4]{Department of Business Intelligence \& Data Science, Parnassia Groep, The Hague, The Netherlands}

\begin{document}

\maketitle

\begin{abstract}

\noindent Classical cluster inference is hampered by the spatial specificity paradox. Given the null-hypothesis of no active voxels, the alternative hypothesis states that there is at least one active voxel in a cluster. Hence, the larger the cluster the less we know about where activation in the cluster is. \cite{Rosenblatt2018} proposed a post-hoc inference method, All-resolutions Inference (ARI), that addresses this paradox by estimating the number of active voxels of any brain region. ARI allows users to choose arbitrary brain regions and returns a simultaneous lower confidence bound of the true discovery proportion (TDP) for each of them, retaining control of the family-wise error rate. ARI does not, however, guide users to regions with high enough TDP. In this paper, we propose an efficient algorithm that outputs all maximal supra-threshold clusters, for which ARI gives a TDP lower confidence bound that is at least a chosen threshold, for any number of thresholds that need not be chosen a priori nor all at once. After a preprocessing step in linearithmic time, the algorithm only takes linear time in the size of its output. We demonstrate the algorithm with an application to two fMRI datasets. For both datasets, we found several clusters whose TDP confidently meets or exceeds a given threshold in less than a second.

\end{abstract}



\section{Introduction}

Cluster inference approaches have become ubiquitous in neuroimaging during the past two decades \citep{Woo2014, Eklund2016}. However, little attention has been paid to the problem of low spatial specificity, especially when the detected significant clusters have a large size \citep{Woo2014, Spisak2019}. \cite{Rosenblatt2018} introduced a novel approach, All-resolutions Inference (ARI), that deals with this problem by giving a simultaneous lower confidence bound for the proportion of truly activated voxels within any chosen cluster, using a closed testing procedure \citep{Goeman2017}. As a post-hoc multiple comparisons approach, ARI allows completely free choice of the rejection region while still controlling the family-wise error rate (FWER) strongly over all statements on activation percentage, i.e.,\ true discovery proportion (TDP), for all possible clusters in the brain. \cite{Weeda2019} has demonstrated that ARI has comparable detection power when compared with random field theory (RFT)-based parametric cluster inference approaches using the NeuroVault datasets \citep{Gorgolewski2015} while always providing strong FWER control. Several permutation-based extensions of TDP-based inference have been proposed \citep{Andreella2020, Vesely2021, Blain2022}.

In cluster inference, a ``cluster" is generally defined as a connected set of voxels. A special type of clusters are \emph{supra-threshold clusters}, maximal clusters with test statistic values all exceeding a given cluster-forming threshold (CFT). The arbitrary choice of the initial CFT has strong influence on the results, and the use of a lenient threshold can lead to clusters with low activation percentage or even false positives \citep{Eklund2016}. Finding the best CFT by optimising the sensitivity while controlling the rate of false positives is non-trivial. The ``threshold-free cluster enhancement (TFCE)" approach \citep{Smith2009} avoids defining a CFT, integrating cluster extent and voxel intensity to provide a TFCE score at each voxel. However, this approach still requires the prespecification of other parameters \citep{Smith2009, Spisak2019}, and does not solve the spatial specificity problem. In view of such problems, a desirable strategy would be to use an adaptive cluster-forming thresholding method and find maximal supra-threshold clusters, each with a different CFT, but all with an activation percentage confidently above a certain threshold, chosen before or after seeing the data. While the original ARI methodology allows this, it does not provide an efficient procedure to find such clusters.

In this paper, we extend ARI by proposing an efficient algorithm that can be used to sequentially generate all desired supra-threshold clusters whose TDP lies above a given threshold with $1-\alpha$ confidence, simultaneously for all possible TDP thresholds between $0$ and $1$. Because these bounds are simultaneous over all thresholds, the final TDP threshold may be chosen post-hoc without compromising FWER control. Based on the new algorithm, researchers can quickly identify regions with a high percentage of activation, but let the desired amount of activation depend on the data. The output clusters are essentially traditional supra-threshold clusters with different CFTs: each with the smallest CFT that guarantees a desired TDP. This approach allows more flexibility in thresholding and offers proper activation localisation. The resulting regions for different TDP thresholds can be visualised simultaneously, enabling a quick assessment of the spatial extent of regions with confidently high percentages of activation. The algorithm takes only $O(m\log m)$ time, where $m$ is the number of voxels under consideration.

We first briefly revisit the theoretical background of ARI in Section \ref{sec: ARI}. The novel adaptive thresholding algorithm is described in detail in Section \ref{sec: algs}, and illustrated with two fMRI applications in Section \ref{sec: apps}. Proofs of the theorems in Section \ref{sec: algs} can be found in the Appendix.

\section{All-resolutions Inference} \label{sec: ARI}

ARI is a cluster inference method proposed by \citet{Rosenblatt2018} to alleviate the spatial specificity problem encountered by traditional cluster inference. It has a more ambitious goal than classical RFT-based cluster inference since it does not only aim to infer the presence of signal in clusters or voxel sets, but also the extent of the activation, measured as the proportion of active voxels (True Discovery Proportion, TDP).

\subsection{Theory of ARI}

Suppose that we are interested in a vertex set $V$ of size $m$, such as a set of voxels related to the whole brain or parts of a brain, where each voxel of the brain is represented by a vertex. Each $v \in V$ has a corresponding null hypothesis $H_v$ to be tested, for which we obtain a $p$-value $p_v$. For simplicity, we identify $V$, without loss of generality, with $[m]$ in such a way that $p_1 \leq p_2 \leq \cdots \leq p_m$, where $[u]$ is the usual shorthand for $\{1, 2, \ldots, u\}$ given an integer $u$. Sorting $p$-values only takes $O(m \log m)$ time and has no impact on the total time complexity of the algorithm presented in Section \ref{sec: algs}.

We denote the set of indices of true null hypotheses by $T \subseteq V$. For every subset $S \subseteq V$, the number of true discoveries (or TDN) in $S$ is $\tau(S) = |S \setminus T|$ and, if $S$ is non-empty, its corresponding activation proportion (or TDP) is denoted by
\[ 
\pi(S) = \tau(S) / |S|.
\]
The TDP informs about the extent of spatial activation within $S$.  Following \cite{Rosenblatt2018}, we say that there is good spatial localisation of the signal in $S$ if the TDP is high enough. 

Based on the results of \citet{Goeman2011} and \citet{Goeman2017}, for $\alpha>0$, ARI offers an $(1-\alpha)$-lower confidence bound $q(S)$ for $\pi(S)$ of all non-empty $S \subseteq V$. It has the property that
\begin{equation*}
\mathbb{P} \big( \pi(S) \geq q(S) \textrm{\ for all non-empty $S \subseteq V$} \big) \geq 1-\alpha.
\end{equation*}
Therefore $[q(S),1]$ is a $(1-\alpha)$-confidence interval for the TDP $\pi(S)$, simultaneous over all possible subsets $S \subseteq V$. The simultaneity of the $(1-\alpha)$-confidence intervals provides a guarantee of post-hoc validity: users may freely choose any rejection sets, in any user-, knowledge- or data-driven ways, while retaining the guarantee that with $1-\alpha$ confidence the proportion of true discoveries in all finally selected sets is not overstated.
Throughout this paper, we will only consider confidence intervals at a fixed confidence level  $1-\alpha$, which we will leave implicit from here on. 

\citet{Rosenblatt2018} explained in detail how to use ARI in the neuroimaging context to find the TDP bounds for voxel sets of interest and to ``drill down'' within such clusters to obtain better spatial localisation of detected signal. Conversely, ARI can also be applied to directly localise activation by creating the largest clusters with the desired amount of true activation, quantified using a TDP threshold. The algorithm described in Section \ref{sec: algs} provides an efficient way to find such clusters.

ARI's error control is guaranteed under the assumption of the Simes inequality which has been proven to hold under independence and for multivariate distributions that satisfy the assumption of positive regression dependency on subsets \citep{Benjamini2001, Sarkar2008b, Su2018}. This assumption is generally satisfied under the typical assumptions made for the fMRI data \citep{Nichols2003}.

\subsection{Current algorithms for ARI} \label{sec: algS1RI}

The TDP lower confidence bound $q(S)$ was derived by \citet{Goeman2017} using the closed testing procedure \citep{Marcus1976} with Simes \citep{Simes1986} local tests. It is given by $q(S) = d(S)/|S|$ for non-empty subsets $S \subseteq V$, where $d(S)$ is the lower confidence bound for the TDN $\tau(S)$ given by
\begin{equation}
d(S) = \begin{cases}
{\max}_{j \in [\lvert S\rvert]}\; \delta(S,j), & \text{if $S \neq \emptyset$,}\\
0, & \text{otherwise,}
\end{cases}
\label{TDP_bound}
\end{equation}
for all $S \subseteq V$,
\[
\delta(S, j) = \bigl\lvert\{v \in S : hp_v \le j\alpha\}\bigr\rvert - j+1
\]
for all $S \subseteq V$ and positive integers $j$,
and
\begin{equation*}
h = \max \big\{ i \in \{ 0,1,\ldots,m \} \colon i p_{m-i+j} > j\alpha~\text{for all}~ j \in [i] \big\}.
\end{equation*}
Since $\delta(S, j) \leq 0 \leq \delta(S, 1)$ for $j > \lvert S\rvert$, 
an equivalent way to express $d(S)$ that will be useful in Section \ref{sec: alg3} is
\begin{equation}
d(S) = {\max}_{j \in [\ell]}\; \delta(S,j)
\label{TDP_bound_extended_range}
\end{equation}
for any positive integer $\ell \geq \lvert S \rvert$.

Calculation of $q(S)$ using ARI is fast for any given $S$. The value of $h$ needs to be calculated only once per data set, and takes linear time in $m$ for sorted $p$-values \citep{Meijer2019}. After that, any $q(S)$ can be calculated in linear time in $|S|$  \citep{Goeman2017}.

Equation \eqref{TDP_bound} is useful and fast if we want to calculate $q(S)$ for a limited number of specific sets of interest. Often, however, a researcher does not have a preconceived idea of what $S$ to look at, but wants to find all clusters that, with confidence, have high enough TDP. Naively trying out all subsets $S$ quickly becomes intractable, so in the rest of this paper we will consider the inverse problem of finding clusters for a given TDP threshold in an efficient manner. In Section \ref{sec: algs}, we will develop an algorithm for finding all maximal supra-threshold clusters among the clusters $S\subseteq V$ with $q(S) \ge \gamma$ for any desired value of $\gamma \in [0,1]$, allowing $\gamma$ to be tuned on basis of the data.

One final practical result from \citet{Goeman2017} we will mention is that the maximum in Equation \eqref{TDP_bound} may be taken over a possibly smaller range.
First, they show that $d(S) = d(S \cap [\zeta])$ for all subsets $S \subseteq V$, where
\begin{equation*}
\zeta =
\begin{cases}
0, & \text{if} ~ h = m, \\
\min \big\{ v \in \{ m-h, \ldots, m \} \colon h p_v \le (v-m+h+1)\alpha \big\}, & \text{otherwise}.
\end{cases}
\end{equation*}
Second, the conditions $hp_v \le j\alpha$ and $c(v) \leq j$ are equivalent for the discretisation 
\begin{equation}
c(v) = \max \bigl\{1, \lceil h p_v / \alpha \rceil \bigr\}
\label{c(v)}
\end{equation}
of $p$-value $p_v$, 
so we have
\[
\delta(S, j) = \bigl\lvert\{v \in S : c(v) \le j\}\bigr\rvert - j+1
\]
for positive integers $j$. 

The implementation of ARI has been provided in the R environment \citep{R} with the R package {\tt hommel} \citep{Goeman2019R}, and specifically for fMRI data analysis, the R package {\tt ARIbrain} \citep{Finos2018R}.

\section{An algorithm for adaptive thresholding} \label{sec: algs}

We suppose our vertex set of interest $V$ is accompanied by a set of edges $E$ to form an undirected graph $G = (V,E)$. For 3-dimensional brain images, there are three conventional ways to specify edges between voxels based on whether two distinct voxels, as geometrical objects, share a voxel face, voxel edge, or voxel vertex, which gives a general voxel 6, 18, or 26 neighbours respectively \citep{Merchant2005,Cheng2009}. In each of these cases the resulting graph is sparse with only $O(m)$ edges.

A \emph{cluster} is a non-empty subset $S \subseteq V$ for which the subgraph of $G$ it induces, denoted by $G[S]$, is connected. A \emph{$\theta$-supra-threshold cluster} is a maximal cluster among all clusters $S \subseteq V$ for which $\max_{v \in S} p_v \leq \theta$, and a \emph{supra-threshold cluster} is a $\theta$-supra-threshold cluster for some threshold $\theta$. The prefix \mbox{``supra-''} comes from the fact that clusters are usually defined in terms of a minimum $z$-score rather than a maximum $p$-value, but the two formulations are equivalent.

In this section we describe an efficient algorithm that answers any number of the following type of queries without knowing them in advance.

\begin{query}
\label{query:maximal-stc}
Given $\gamma \in [0,1]$, find all maximal supra-threshold clusters among the supra-threshold clusters $S \subseteq V$ for which $q(S) \geq \gamma$.
\end{query}

The algorithm builds a data structure to handle these queries in three steps. The first step finds all possible supra-threshold clusters in $O(\lvert E\rvert a(m))$ time and $O(m)$ space, where $\lvert E\rvert$ is the number of edges in the graph, and $a(m)$ is the inverse of the diagonal Ackermann function that grows extremely more slowly than $\log(m)$. The second step calculates the TDP lower confidence bounds for all these supra-threshold clusters in $O(m \log m)$ time and $O(m)$ space. The third step takes $O(m \log m)$ time and $O(m)$ space to filter out unfeasible supra-threshold clusters that cannot satisfy any query, and to sort the remaining supra-threshold clusters on their TDP lower confidence bounds. Together, that makes $O(\lvert E\rvert a(m) + m\log m)$ time and $O(m)$ space to create our data structure, which becomes $O(m \log m)$ time and $O(m)$ space for the cases we have in mind where $G$ is sparse with only $O(m)$ edges. After the construction is completed, any query can be answered in linear time in the sum of the sizes of the maximal supra-threshold clusters that satisfy the query, i.e., answering queries is output-sensitive.

It follows from the next subsection that the maximal supra-threshold clusters satisfying a query are pairwise disjoint (and the subgraphs of $G$ they induce pairwise non-adjacent).

\subsection{Finding all supra-threshold clusters} \label{sec: cluster indexing}

We start with enumerating all supra-threshold clusters, without regard to their TDP lower confidence bounds. We remark that the $\theta$-supra-threshold clusters are exactly the vertex sets of the (connected) components of the subgraph of $G$ induced by the vertices $\{v \in V : p_v \leq \theta\}$. Moreover, we may assume the threshold $\theta$ is a $p$-value without loss of generality. That means the $p_u$-supra-threshold clusters are the vertex sets of the components of the induced subgraph $G[[u]]$, where $u \in V$ is the largest vertex with $p$-value $p_u$. In the typical situation where all $p$-values are distinct, this amounts to determining the components of $G[[v]]$ for every vertex $v \in V$. In general, we will encounter the components of each subgraph $G[[v]]$ naturally while computing the supra-threshold clusters. 

Fortunately, there are only $m$ distinct such components in total. Each induced subgraph $G[[v]]$ has a unique component $C_v$ containing $v$, and any other component $C$ of $G[[v]]$ is the unique component of $G[[u]]$ and is therefore identical to $C_u$, where $u < v$ is the largest vertex in $C$.

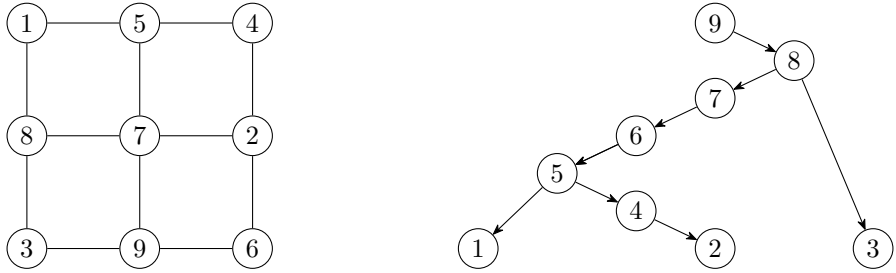
\begin{figure}
\centering
\begin{tikzpicture}[every node/.style={circle,draw, inner sep=0pt, minimum size=15pt}, scale=1.5]
    \node(A) at (0,2) {$1$}; 
    \node(B) at (1,2) {$5$}; 
    \node(C) at (2,2) {$4$}; 
    \node(D) at (0,1) {$8$}; 
    \node(E) at (1,1) {$7$}; 
    \node(F) at (2,1) {$2$}; 
    \node(G) at (0,0) {$3$}; 
    \node(H) at (1,0) {$9$}; 
    \node(I) at (2,0) {$6$}; 
    
    \path (A) edge (B) edge (D);
    \path (E) edge (B) edge (D) edge (F) edge (H);
    \path (G) edge (D) edge (H);
    \path (I) edge (H) edge (F);
    \path (C) edge (B) edge (F);
  
    \begin{scope}[xshift=4cm, yscale=1/3, xscale=.7, >={Stealth[round]}]
        \node(1) at (0,0) {$1$}; 
        \node(2) at (3,0) {$2$}; 
        \node(3) at (5,0) {$3$};
        \node(4) at (2,1) {$4$}; 
        \node(5) at (1,2) {$5$}; 
        \node(6) at (2,3) {$6$};
        \node(7) at (3,4) {$7$};
        \node(8) at (4,5) {$8$};
        \node(9) at (3,6) {$9$}; 
        
        \path[->] (9) edge (8); 
        \path[->] (8) edge (7) edge (3);
        \path[->] (7) edge (6);
        \path[->] (6) edge (5);        
        \path[->] (6) edge (5);        
        \path[->] (5) edge (1) edge (4);
        \path[->] (4) edge (2);        
    \end{scope}
\end{tikzpicture}
\caption{Illustration of input and output of Algorithm \ref{algA}. On the left we have a sparse, undirected graph $G$ that corresponds to nine voxels that are connected by edges through shared voxel faces. Vertices are numbered in ascending order of their $p$-values. On the right we have the directed rooted forest (in this case a tree because $G$ is connected) found by Algorithm \ref{algA}. 
We see that, for instance, the two components of the induced subgraph $G[[5]]$ have vertex sets $\{3\}$ and $\{1,2,4,5\}$, which correspond to the subtrees with roots $3$ and $5$ on the right respectively. Note also that the edge between $6$ and $2$ on the left gives rise to the edge from $6$ to the representative $5$ of $\{1,2,4,5\}$ on the right.}
\label{fig:disjoint_forest}
\end{figure}

We may neatly and compactly organize the vertex sets of the components $(C_v)_{v \in V}$ in a directed rooted forest $\mathcal F$ as follows. We appoint $v$ as the representative of $C_v$, and ensure that the subtree of $\mathcal F$ rooted at $v$ contains precisely the vertices in $C_v$. Since $C_v$ consists of $v$ and the vertices of all preceding $C_u~(u < v)$ adjacent to $v$, all we need to do to add vertex $v$ to $\mathcal F$ when the forest already contains all vertices less than $v$ is to add an edge from $v$ to these $u$.
Algorithm \ref{algA} formalises how to construct $\mathcal F$, and Figure \ref{fig:disjoint_forest} shows the output of this algorithm on a sample input. The supra-threshold clusters are easily distinguished as the vertex sets of those components $C_v$ for which $v$ has no parent $u$ in $\mathcal F$ with $p_u = p_v$. 

\begin{algorithm}[t]
\caption{Compute a forest representation of the vertex sets of the components $(C_v)_{v \in V}$.}
\label{algA}
\begin{algorithmic}[0]
\Require{$p_1 \leq p_2 \leq \cdots \leq p_m$}
\Statex
\Function{FindClusters}{$V$, $E$}
	\State Initialise an edgeless directed rooted forest $\mathcal F$ with vertices $V$.
    \Statex
	\For{$v = 1$ \To $m$}
		\ForAll{$\{u,v\} \in E$ such that $u < v$}
				\State Find the root $w$ of the subtree in $\mathcal F$ that contains $u$.
				\If{$v \neq w$}
			 		\State Add edge $(v, w)$ to $\mathcal F$.
				\EndIf
		\EndFor
	\EndFor
	\Statex
	\State \Return{$\mathcal{F}$}  
\EndFunction
\end{algorithmic}
\end{algorithm}

The only step in Algorithm \ref{algA} that is non-trivial to implement is finding the root of the subtree in $\mathcal F$ that contains a given vertex. In order to quickly locate these roots, we keep track, as edges are added to $\mathcal F$, of the root of each component of $\mathcal F$ and to which component of $\mathcal F$ each vertex belongs. This is an instance of the incremental connectivity problem, a well-known problem in graph theory that has an efficient solution using a disjoint-set data structure \citep{Tarjan1975}. Performing $k$ find and union operations on a set of $m$ elements takes $O(ka(m))$ time with this data structure \citep{Tarjan1975}. Since Algorithm \ref{algA} performs one find and at most one union operation for each edge of $G$, the total time complexity of Algorithm \ref{algA} becomes $O(\lvert E \rvert a(m))$. Storage is linear in $m$.

\subsection{Calculating TDP lower confidence bounds for an ascending chain of subsets} \label{sec: alg3}

From the result of Algorithm \ref{algA} and Equation \eqref{TDP_bound} we could calculate the TDP lower confidence bounds for all supra-threshold clusters in $O(m^2)$ time. We will be able to do better, however, by exploiting an algorithm proposed by \cite{Meijer2015} to calculate TDP lower confidence bounds efficiently for an entire ascending chain at once. 

\begin{figure}
\centering
\begin{tikzpicture} [scale=1.5]
\datavisualization [
	scientific axes=clean, 
	x axis={label={$k$}},
	y axis={label={$f_i(k)$}},
	visualize as line/.list={f0,f1,f2,f3,f4}, 
	every visualizer/.style={mark=*, mark size=1.4pt, thin, gap around stream point=3pt}, 
	f0={gap line, label in data={text={$f_0$}, index=5, node style={right}}},
	f1={gap line, label in data={text={$f_1$}, index=5, node style={right}}},
	f2={gap line, label in data={text={$f_2$}, index=5, node style={right}}},
	f3={gap line, label in data={text={$f_3$}, index=5, node style={right}}},
	f4={gap line, label in data={text={$f_4, f_5$}, index=5, node style={right}}},
	visualize as scatter=categories,
	categories={style={mark=o, mark size=3pt}}
]
data [set=f0] {
x, y
1, 0
2, -1
3, -2
4, -3
5, -4
}
data [set=f1] {
x, y
1, 0
2, -1
3, -1
4, -2
5, -3
}
data [set=f2] {
x, y
1, 1
2, 0
3, 0
4, -1
5, -2
}
data [set=f3] {
x, y
1, 1
2, 0
3, 0
4, -1
5, -1
}
data [set=f4] {
x, y
1, 1
2, 1
3, 1
4, 0
5, 0
}
data [set=categories] {
x, y
3, -1
5, -1
1, 1
3, 1
};
\end{tikzpicture}
\caption{Sequences $f_0, f_1, \ldots, f_5$ for an ascending chain with defining sequence $(v_i)_{i=1}^5$ such that $\bigl(c(v_i)\bigr)_{i=1}^5 = (3,1,5,3,6)$ are its discretised $p$-values. The points that belong to each $f_i$ lie on the unique curve that ends in the point labelled by $f_i$. Except for $f_5$, the index of the rightmost circled point of each $f_i$ indicates the discretised $p$-value $c(v_i)$.
We may observe that the pointwise difference $f_{i+1} - f_i$ of consecutive sequences increases weakly from 0 to 1. Note that the update from $f_3$ to $f_4$ not only forces changes on $k \geq c(v_4) = 3$, but also on $k=2$.}
\label{fig:chain_TDP_bounds}
\end{figure}
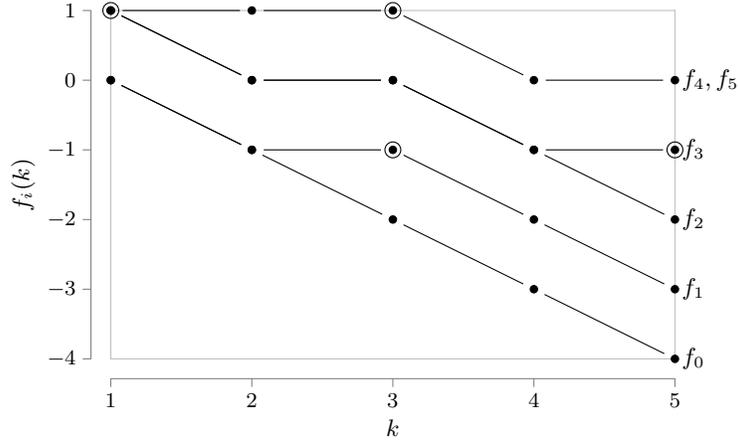

Let us restate the algorithm of \cite{Meijer2015} together with its correctness argument. 
Given an ascending chain of subsets
\[\emptyset = V_0 \subset V_1 \subset \cdots \subset V_{\ell} \subseteq V\]
with $\lvert V_{\ell} \rvert = \ell$, let $(v_i)_{i=1}^{\ell}$ be its defining sequence given by $V_i\setminus V_{i-1}= \{v_i\}$. For each subset $V_i$ in this chain, consider the weakly decreasing integer sequence $\bigl(f_i(k)\bigr)_{k=1}^{\ell}$ given by 
\[f_i(k) = \max_{j \in \{k,k+1,\ldots,\ell\}} \delta(V_i, j) \]
that, in reverse order, calculates Equation \eqref{TDP_bound_extended_range} for $S = V_i$ by taking the maximum in that equation over more and more terms until we eventually get to $f_i(1) = d(V_i)$. See Figure \ref{fig:chain_TDP_bounds} for an example. Since the difference between distinct, consecutive elements of $f_i$ must always be $-1$, it follows that $f_i$ is completely determined by the value of $f_i(1)$ and its fibers, i.e., the parts of its domain on which it is constant.

Algorithm \ref{algB} computes this representation of $f_i$ from the representation of $f_{i-1}$ to calculate the lower confidence bound $d(V_i)$ of each non-empty subset $V_i$ in the chain, from which $q(V_i)$ immediately follows. The correctness of Algorithm \ref{algB} is established by Theorem~\ref{theorem_1}.

\begin{algorithm}[t]
\caption{Calculate TDN lower confidence bounds of subsets that form an ascending chain.}
\label{algB}
\begin{algorithmic}[0]
\Require{$c_1 = c(v_1),\, c_2 = c(v_2),\, \ldots,\, c_{\ell} = c(v_\ell)$ for some distinct $v_1, v_2, \ldots, v_{\ell} \in V$}
\Statex
\Function{ComputeTDNBounds}{$c_1, c_2, \ldots, c_{\ell}$}
	\Let{$d_0$}{$0$}	\Comment{$d_i$ tracks $f_i(1)$}
	\Let{$\mathcal D$}{$\big\{ \{1\}, \{2\}, \ldots, \{\ell \} \big\}$}	\Comment{$\mathcal D$ tracks the non-empty fibers of $f_i$}
	\Statex
	\For{$i = 1$\To $\ell$}
		\Let{$d_i$}{$d_{i-1}$}
		\If{$c_i \leq \ell$}
			\State Find $I \in \mathcal D$ such that $c_i \in I$.
			\If{$\min I = 1$}
			 	\Let{$d_i$}{$d_i + 1$}
			\Else
				\State Find $J \in \mathcal D$ such that $\min I - 1 \in J$.
				\Let{$\mathcal D$}{$\bigl(\mathcal D \setminus \{I,J\}\bigr) \cup \{I \cup J\}$}
			\EndIf
		\EndIf
	\EndFor
	\Statex
	\State \Return{$(d_1, d_2, \ldots, d_{\ell})$}
\EndFunction
\end{algorithmic}
\end{algorithm}


\begin{restatable}{theorem}{algBCorrectness} \label{theorem_1}
Algorithm \ref{algB} returns $\bigl(d(V_i)\bigr)_{i=1}^{\ell}$ when given discretised $p$-values $\bigl(c(v_i)\bigr)_{i=1}^{\ell}$, where $(v_i)_{i=1}^{\ell}$ is the defining sequence of an ascending chain of subsets $\emptyset = V_0 \subset V_1 \subset \cdots \subset V_{\ell} \subseteq V$ with $\lvert V_{\ell}\rvert = \ell$.
\end{restatable}

An implementation of Algorithm \ref{algB} is straightforward except for a choice of data structure for $\mathcal D$. The set $\mathcal D$ partitions $[\ell]$ into consecutive integer intervals, because it does so initially and each iteration of the loop in Algorithm \ref{algB} maintains that invariant. A data structure for $\mathcal D$ needs to support queries to which interval a discretised $p$-value belongs and be able to merge an interval with the interval that comes directly before it (if any), so a natural choice is a disjoint-set data structure. We augment this data structure to keep track of the smallest element of each interval, so we do not need to recompute that element each time it is requested. Since we only merge adjacent intervals instead of arbitrary sets, a slightly more efficient implementation of a disjoint-set data structure is available that makes Algorithm \ref{algB} run in $O(\ell)$ time rather than $O(\ell a(\ell))$ time \citep{Gabow1985}. In practice, this theoretically improved version appears to be competitive with the simpler, usual version of a disjoint-set data structure \citep{Gabow1985}. Storage of Algorithm \ref{algB} is also $O(\ell)$. A further practical improvement to Algorithm \ref{algB} is the use of the identity $d(V_i) = d(V_i \cap [\zeta])$ from Section \ref{sec: algS1RI} to shrink $V_i$.

\subsection{Calculating TDP lower confidence bounds for all supra-threshold clusters} \label{sec: forest TDP}

Now suppose we constructed the directed rooted forest $\mathcal F$ from $G$ as in Section \ref{sec: cluster indexing}. Recall that a (\emph{directed}) \emph{path} of $\mathcal F$ is a non-empty sequence $(v_i)_{i=1}^k$ of vertices of $\mathcal F$ such that each pair $(v_i, v_{i+1})$ is an edge of $\mathcal F$, and that a \emph{path cover} of $\mathcal F$ is a set of paths that together contain every vertex of $\mathcal F$. The basic idea to compute TDP lower confidence bounds for each cluster encoded in $\mathcal F$ is as follows. 

We first pick a path cover $\mathcal P$ of $\mathcal F$. For each path $P \in \mathcal P$, we traverse the subtree of $\mathcal F$ rooted at the start of $P$ in post-order, where we prioritise visiting a child that lies on $P$ over any of the child's siblings. While doing so, we list all vertices of that subtree in a sequence $L_P^{}$, called the \emph{sequentialisation} of that subtree, in the order that we encounter them during this traversal.  By ordering vertices in this way, the first terms of $L_P^{}$ up to and including a vertex $v$ on $P$ are precisely the vertices in the subtree of $\mathcal F$ rooted at $v$, that is, the vertices in the component $C_v$. We then run Algorithm \ref{algB} on the discretised $p$-values of the sequentialisation $L_P^{}$ to find the TDP lower confidence bounds for the vertex set of $C_v$ for each vertex $v$ that lies on $P$.

After choosing a path cover of $\mathcal F$, computing all TDP lower confidence bounds in this manner takes $O\bigl(\sigma(\mathcal F, \mathcal P)\bigr)$ time and $O(m)$ space. Here, we define $\sigma(\mathcal F, \mathcal P) = \sum_{v \in I(\mathcal P)} \lvert\mathcal F_v \rvert$, where $I(\mathcal P)$ is the set of starting vertices of the paths in $\mathcal P$, $\mathcal F_v$ is the subtree of $\mathcal F$ rooted at $v$, and $\lvert \mathcal F_v\rvert$ is the order of $\mathcal F_v$ (i.e., its number of vertices).

Take for example the path cover $\tilde{\mathcal P}$ of the graph in Figure \ref{fig:disjoint_forest} by the paths from $4$ to $2$, from $7$ to $1$, and from $9$ to $3$. Post-order traversals of the associated forest $\tilde{\mathcal F}$ with respect to each of these paths yield the sequentialisations $(2,4)$, $(1,2,4,5,6,7)$, and $(3,1,2,4,5,6,7,8,9)$ respectively. Note that $3$ has priority over $7$ in such a traversal for the last path, so $3$ occurs before vertices in the subtree rooted at $7$ in the last sequentialisation. 
On the other hand, neither $1$ has priority over $4$ in that traversal nor vice versa, so another valid sequentialisation would be $(3,2,4,1,5,6,7,8,9)$. No matter which sequentialisation is settled upon for the last path, only its first terms up to and including $3$, $8$, and $9$ correspond to the vertices in a rooted subtree of $\tilde{\mathcal F}$. The sum of the lengths of the sequentialisations is $\sigma(\tilde{\mathcal F}, \tilde{\mathcal P}) = 17$ for this path cover.

We regard a path cover of $\mathcal F$ to be \emph{optimal} if it minimises $\sigma(\mathcal F, \mathcal P)$ over all path covers $\mathcal P$ of $\mathcal F$, and we denote $\sigma(\mathcal F)$ for the minimum value it attains. The following theorem characterises optimal path covers, and asserts that their paths are picked in a greedy fashion. Here, we say an edge $(u,v)$ in $\mathcal F$ is \emph{heavy} if $\lvert \mathcal F_v\rvert \geq \lvert \mathcal F_w\rvert$ for each child $w$ of $u$. A path is called \emph{heavy} if it only takes heavy edges, and a \emph{heavy} path cover of $\mathcal F$ is a path cover of $\mathcal F$ by heavy paths. Recall further that a path cover of $\mathcal F$ is \emph{minimal} if it has the least number of paths among all path covers of $\mathcal F$, and \emph{vertex-disjoint} if it does not have two distinct paths that share a vertex. The forest in Figure \ref{fig:disjoint_forest} has a unique minimal, vertex-disjoint, heavy path cover $\tilde{\mathcal P}'$ by the heavy paths from $9$ to $2$, from $1$ to $1$, and from $3$ to $3$, for which the lengths of the sequentialisations sum to $\sigma(\tilde{\mathcal F}, \tilde{\mathcal P}') = 11$.

\begin{restatable}{theorem}{optimalpathcovers} \label{theorem_2}
A path cover $\mathcal P$ of $\mathcal F$ is optimal if and only if it is minimal, vertex-disjoint, and heavy.
\end{restatable}

Observe that the set of paths from each root of $\mathcal F$ to each leaf of $\mathcal F$ is a minimal path cover, so a path cover of $\mathcal F$ is minimal precisely when 
each of its paths ends in a leaf of $\mathcal F$ and distinct paths end in distinct leaves of $\mathcal F$. Furthermore, a minimal, vertex-disjoint path cover uses exactly one outgoing edge from every non-leaf of $\mathcal F$, and, conversely, any fixed choice of outgoing edge from each non-leaf of $\mathcal F$ gives rise to such a path cover.
That means it is easy to construct an optimal path cover $\mathcal P$ of $\mathcal F$ while building the directed rooted forest $\mathcal F$ itself bottom-up. We simply modify Algorithm~\ref{algA} to maintain a selection of a single outgoing heavy edge for each non-leaf of $\mathcal F$ while edges are added to $\mathcal F$, for which we also need to keep track of the order of each rooted subtree. This modification does not affect the time or space complexity of Algorithm~\ref{algA}.

In an implementation there is no need to mark the heavy edges of $\mathcal F$ taken by paths in $\mathcal P$ explicitly; we may convey this information by putting the child that a chosen heavy edge points to at the front of the list of children it belongs to. This technique has the added benefit that performing post-order traversals of subtrees of $\mathcal F$ by iterating over the children of a vertex in order automatically gives priority to edges of a path in $\mathcal P$.

The following theorem estimates $\sigma(\mathcal F)$ in the worst case. We denote $\sigma(m) = \max_{\mathcal F'} \sigma(\mathcal F')$, where the maximum runs over all directed rooted forests $\mathcal F'$ of order $m$.

\begin{restatable}{theorem}{worstcasesigma} \label{theorem_3}
We have $\sigma(m) = m \log_4 m + O(m)$ for each positive integer $m$.
\end{restatable}

The preceding theorem implies that all TDP lower confidence bounds can be computed as described above in $O(m \log m)$ time. We note that path covers need to be chosen carefully to achieve such a time complexity, because an unbalanced caterpillar tree $\mathcal F$ of the form below
\begin{center}
\begin{tikzpicture}[every node/.style={circle,draw, inner sep=0pt, minimum size=23pt}, scale=1.6, >={Stealth[round]}]
	\node(1) at (0,0) {\small$1$};
	\node(2) at (1,0) {\small$2$};
	\node(4) at (3,0) {\small$\frac{m}{2}$}; 
	\node(5) at (3,1) {\small$\frac{m}{2}\mathord+1$};
	\node[draw=none](6) at (2,1) {$\;\ldots\;$};
	\node(7) at (1,1) {\small$m\mathord-1$};
	\node(8) at (0,1) {\small$m$};

	\draw[->] (8) edge (7) edge (1);
	\draw[->] (7) edge (6) edge (2);
	\draw[->] (6) edge (5);
	\draw[->] (5) edge (4);
\end{tikzpicture}
\end{center}
admits a minimal, vertex-disjoint path cover $\mathcal P$ whose paths are given by the vertical edges, in which case $\sigma(\mathcal F, \mathcal P) = \sum_{i=1}^{m/2} 2i = \Theta(m^2)$.

\begin{remark}
For certain families of directed rooted forests $\mathcal{F}$ we even have that $\sigma(\mathcal{F})$ is linear in the order of $\mathcal{F}$. 
%
E.g., it follows from Theorem~\ref{theorem_3} that $\sigma(\mathcal F) = O(m)$ for each directed rooted forest $\mathcal F$ with a path through all but $O(m/ \log m)$ of its $m$ vertices.
%
The forests generated by the applications in Section \ref{sec: apps} could be considered members of this family, as their longest path comprises 86\% and 87\% of their total of 225212 and 145872 vertices, respectively.
\end{remark}

\subsection{Querying for maximal supra-threshold clusters with sufficient TDP lower confidence bounds} \label{sec: query}

Given the directed rooted forest $\mathcal F$ that encodes the supra-threshold clusters of the undirected graph $G$ as in Section \ref{sec: cluster indexing}, together with a TDP lower confidence bound for each of these clusters, answering Query~\ref{query:maximal-stc} for a specified $\gamma \in [0,1]$ is a simple matter of traversing $\mathcal F$ starting from its roots. Whenever we visit a vertex $v$ during this traversal, we output the subtree of $\mathcal F$ rooted at $v$ if the vertex set $S_v$ of $C_v$ is a supra-threshold cluster (checked using the condition in Section \ref{sec: cluster indexing}) and $q(S_v) \geq \gamma$, and otherwise we recursively traverse each child of $v$ in search of qualifying supra-threshold clusters. Processing a query in such manner always takes $\Theta(m)$ time.

It is possible to do a little better in general cases after an additional preprocessing step. 
Let us call a vertex $v$ of $\mathcal F$ \emph{admissible} if $S_v$ is a supra-threshold cluster of $G$ and $q(S_v) > q(S_u)$ for each ancestor $u$ of $v$ in $\mathcal F$. We first collect all admissible vertices of $\mathcal F$ in a list $L$ using a straightforward traversal of $\mathcal F$ that keeps track of the maximum of $q(S_u)$ over all ancestors $u$ of a visited vertex. We then sort the admissible vertices $v$ in $L$ in ascending order of $q(S_v)$. This preprocessing takes $O(m \log m)$ time and $O(m)$ space in total.
We also reserve an additional $O(m)$ space for marking admissible vertices when handling queries.

Resolving Query~\ref{query:maximal-stc} for a given $\gamma \in [0,1]$ can now be done as follows. We first run a linear search of $L$ for its left-most vertex $u$ for which $q(S_u) \geq \gamma$, starting from the last vertex $w$ of $L$. For each unmarked vertex $v$ in $L$ from $u$ to $w$, we then output $S_v$ by traversing the subtree of $\mathcal F$ rooted at $v$ and we mark each vertex we output. Afterwards, we clear all marks to be ready for potential subsequent queries. Answering a query this way takes linear time in the size of the output clusters. We remark that the search part can be improved in practice by running a binary search for the same vertex $u$ in parallel with the linear search, stopping both searches when either is successful.

\subsection{Implementation}

The adaptive thresholding algorithm has been implemented in the R package {\tt ARIbrain} that is available on GitHub \citep{Finos2022}.

We measured the running time of the adaptive algorithm for two families of graphs with artificial $p$-values and $\alpha=0.05$. The first family consists of the graphs, for $m \in \{10^3, 25^3, \ldots, 205^3\}$, that arise by taking $m$ voxels arranged in a cube as vertices, and defining an edge between each pair of distinct vertices that have a voxel edge in common. The second family comprises the perfect binary trees of order $m \in \{2^{10}-1, 2^{11}-1, \ldots, 2^{23}-1\}$, which, we recall, are rooted trees where every non-leaf has two children and the paths from root to leaves are all equally long. The artificial $p$-values for the vertices of the graphs in both families are independently drawn from the cube of the standard uniform distribution, except that for perfect binary trees $G$ the generated $p$-values are permuted in such a way that $G$ is isomorphic to the directed rooted forest $\mathcal F$ from Section~\ref{sec: cluster indexing} if we ignore directionality of edges.

\begin{figure}
\centering
\subfloat[Data structure construction for cubes of voxels.]{\label{fig_timing_1}\includegraphics[width=0.6\textwidth]{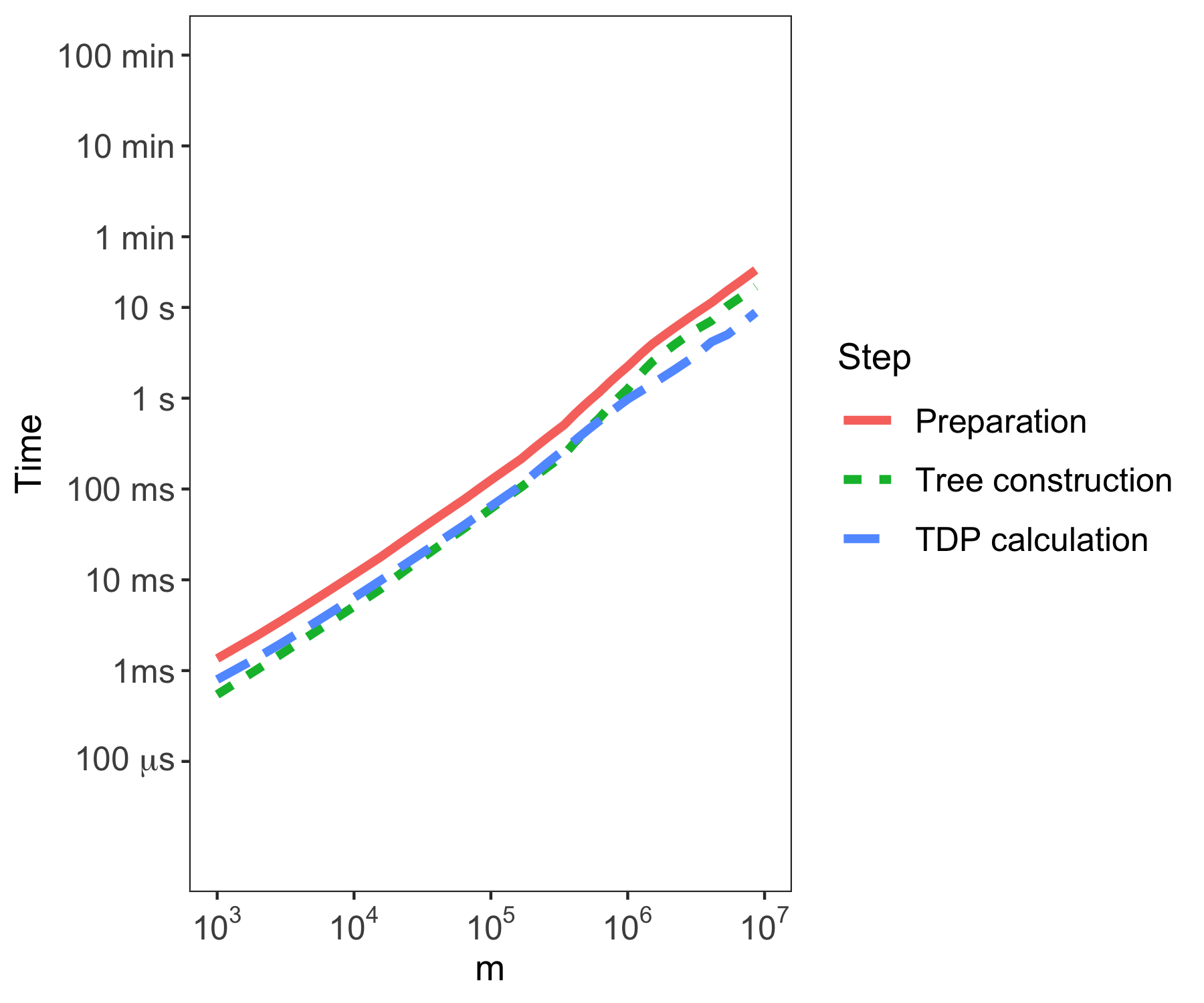}}
\subfloat[Query handling for cubes of voxels.]{\label{fig_timing_2}\includegraphics[width=0.4\textwidth]{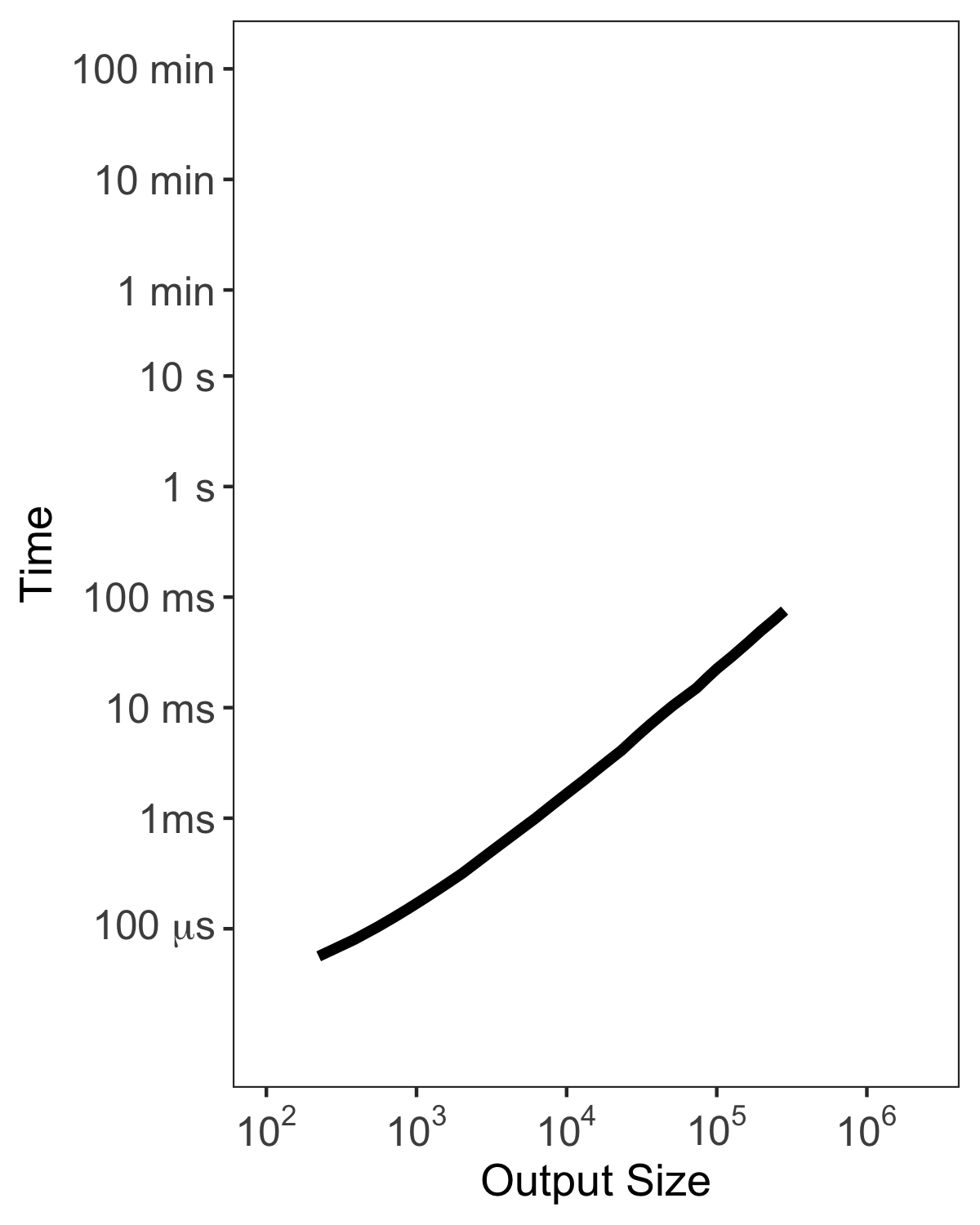}}
\par\medskip
\centering
\subfloat[Data structure construction for perfect binary trees.]{\label{fig_timing_3}\includegraphics[width=0.6\textwidth]{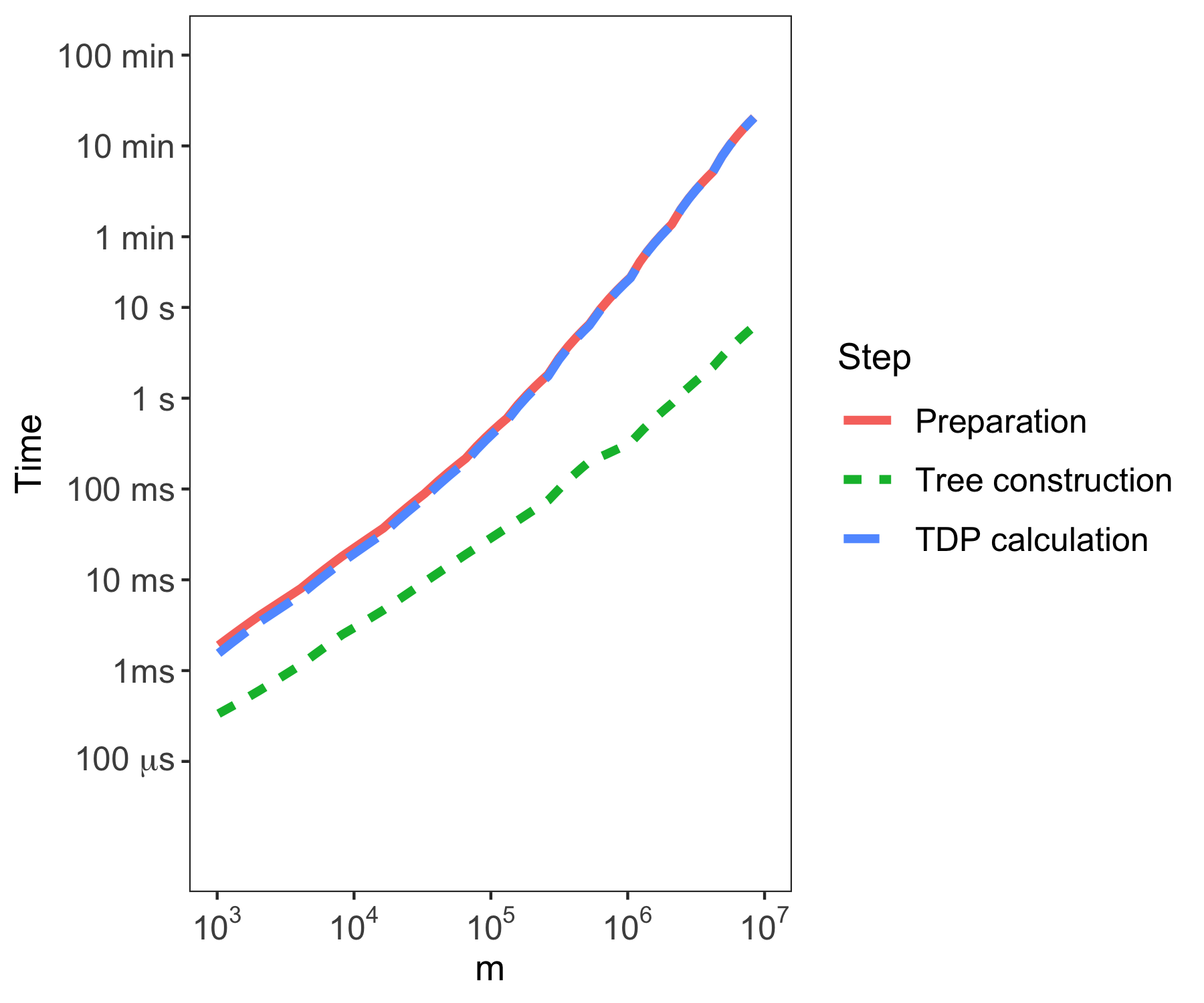}}
\subfloat[Query handling for perfect binary trees.]{\label{fig_timing_4}\includegraphics[width=0.4\textwidth]{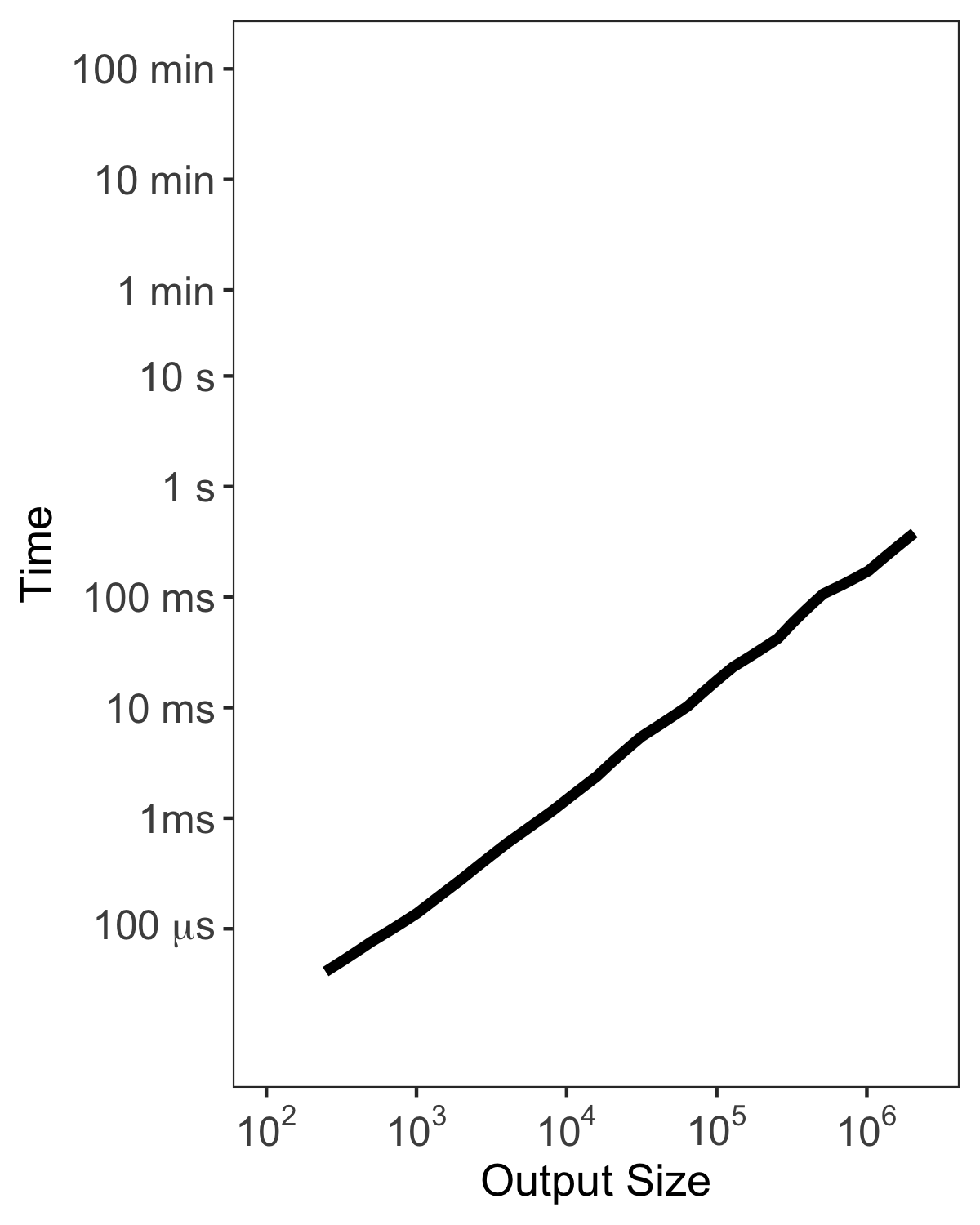}}
\caption{
Average running times on a log-log scale of the adaptive thresholding algorithm, which constructs a data structure (see left figures) to handle Query~\ref{query:maximal-stc} for various TDP thresholds (see right figures), on a family of graphs corresponding to cubes of voxels (see top figures) and a family of perfect binary trees (see bottom figures). 
The figures on the left additionally show the contribution to the total average running time of the three steps described in Section~\ref{sec: cluster indexing}, Section~\ref{sec: forest TDP}, and Section~\ref{sec: query}.
}
\label{fig: timing}
\end{figure}

We ran the adaptive algorithm 100 times on all graphs in both families with different artificial $p$-values each time, and measured its average running time to construct its data structure as detailed in Section~\ref{sec: cluster indexing} through Section~\ref{sec: query}. Furthermore, for all constructed graphs in each family we also measured the average running time to resolve Query~\ref{query:maximal-stc} for TDP thresholds ranging within $\{0, 0.01, \ldots, 1\}$. The resulting timings can be seen in Figure~\ref{fig: timing}. We observe that all steps took almost $O(m)$ time except the preparation step of TDP calculation for the perfect binary trees (see Figure \ref{fig_timing_3}), which has a linearithmic time complexity. Comparing Figures \ref{fig_timing_1} \& \ref{fig_timing_3}, we notice that, preparation steps of tree construction and TDP calculation have similar running time for the first family, while for the second family (the constructed worst-case scenario), the TDP calculation step is dominant as expected. We also note that the asymptotic running time of the query stage linearly depends on the output size (see Figures \ref{fig_timing_2} \& \ref{fig_timing_4}), matching its linear time complexity.

\section{Applications} \label{sec: apps}

We illustrate the use of the novel algorithm by analysing two fMRI datasets that were previously analysed with ARI by \citet{Rosenblatt2018}, specifying the graph $G$ by constructing edges between distinct voxels with common voxel edges, and choosing $\alpha = 0.05$.

\subsection{Go/No-go dataset}

The Go/No-go dataset consists of 34 subjects. Each subject completed an emotional Go/No-go task which required them to press a button under a ``Go'' condition (seeing emotional faces) but not under a ``No-go'' condition (seeing neutral faces). 
The functional brain images of the subjects were acquired and analysed using FSL \citep{Smith2004,Jenkinson2012}, as described in the original publication \citep{Lee2018}. The No-go$>$Go contrast was computed using FSL FEAT \citep{Woolrich2001} for group-level analysis. 

\begin{figure}
\begin{center}
\includegraphics[trim = 0mm 0mm 0mm 0mm, clip=true, width=\textwidth]{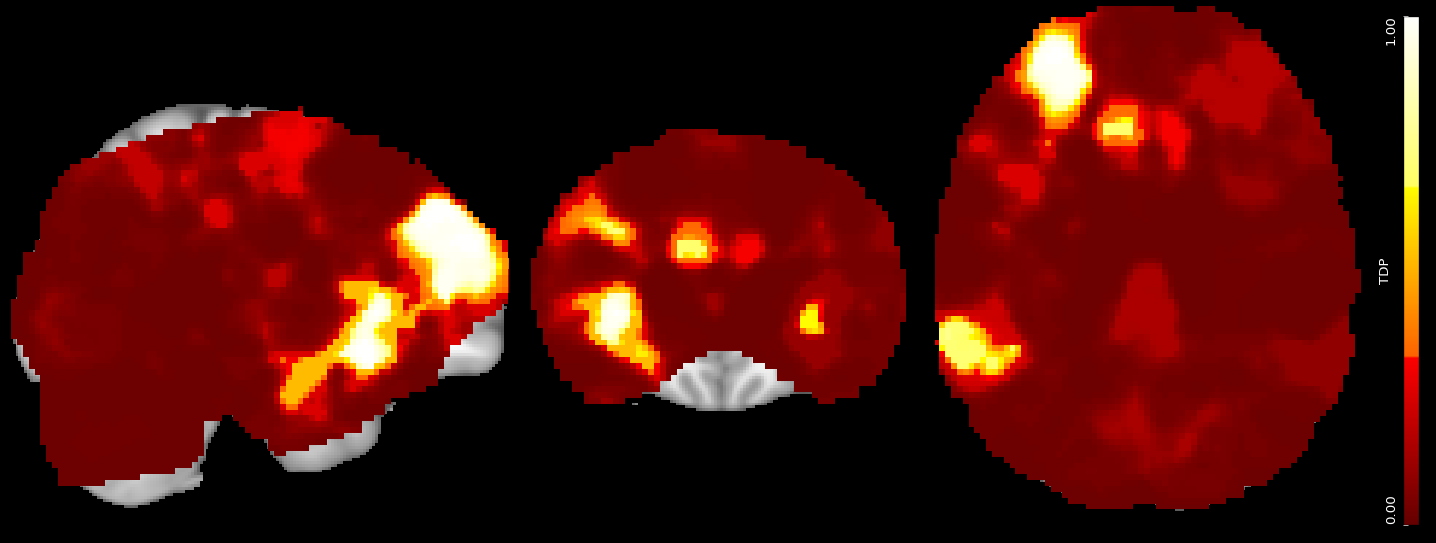}
\caption{A heat map for the Go/No-go dataset that shows for each voxel $v$ the maximum TDP threshold $\gamma_v$ for which $v$ appears in a supra-threshold cluster $S$ with $q(S) \geq \gamma_v$. Brighter colours correspond to higher TDP thresholds.}
\label{fig: grad_gonogo}
\end{center}
\end{figure}

\begin{figure}[ht]
\begin{center}
\includegraphics[trim = 0mm 0mm 0mm 0mm, clip=true, width=\textwidth]{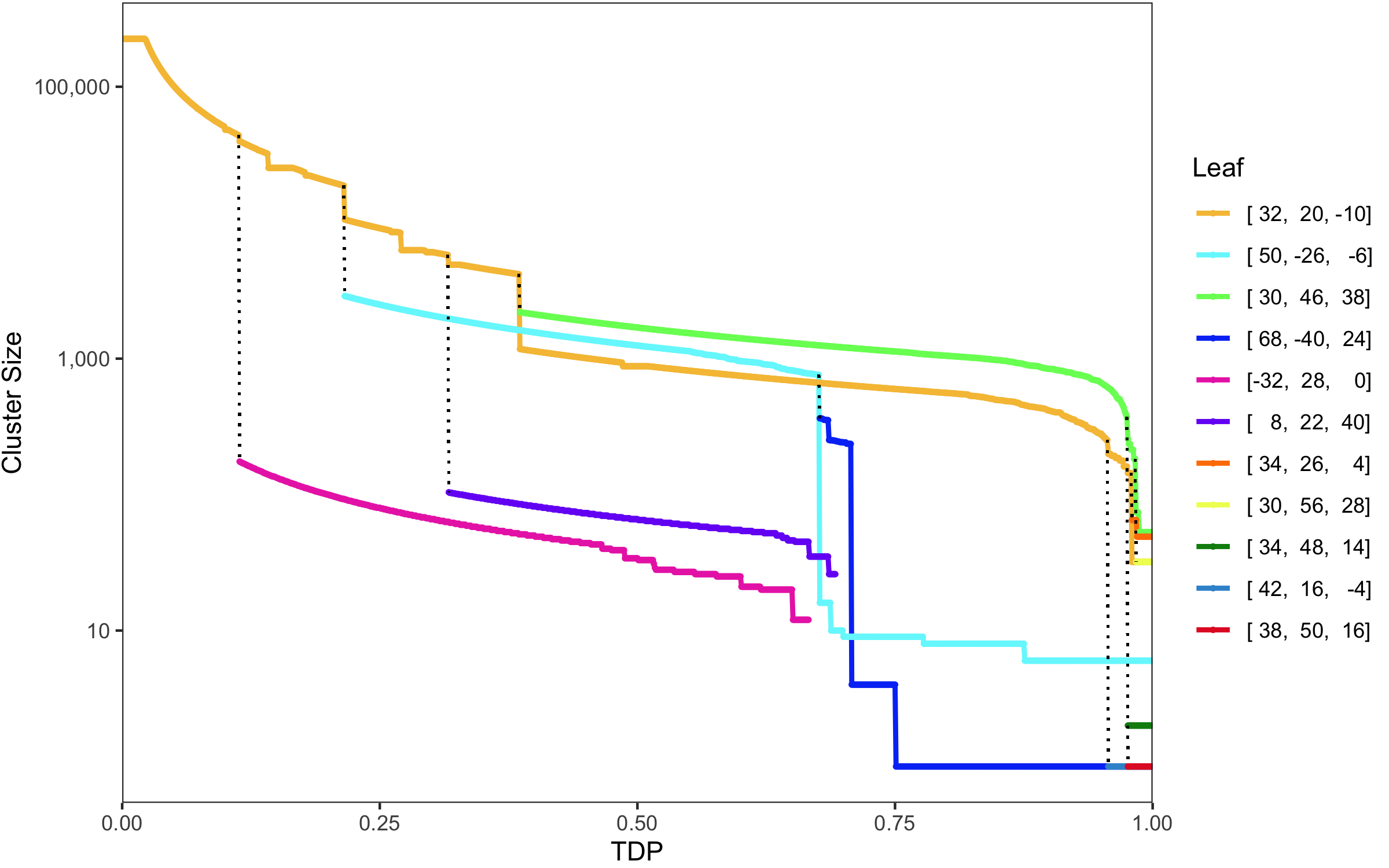}
\caption{A line plot for the Go/No-go dataset that shows how the size of adaptive maximal supra-threshold clusters (on a log scale) varies with the TDP threshold. Clusters are categorised by the MNI coordinates of the end of the heavy path through the cluster's largest vertex. Dashed lines indicate cluster separation, where a cluster splits into the subclusters connected by the dashed line for more stringent TDP thresholds.}
\label{fig: tks_gonogo}
\end{center}
\end{figure}

After the requisite preprocessing was done, it took around 0.3 seconds to build the adaptive thresholding data structure for $m=225212$ in-mask voxels.
Resolving a query for maximal supra-threshold clusters for each TDP threshold $\gamma \in \{0, 0.01, 0.02, \ldots, 1\}$ from this data structure took about 0.003 seconds on average for an average output size of 19849 voxels. Figure \ref{fig: grad_gonogo} visualises the maximal supra-threshold clusters at all TDP thresholds simultaneously. Figure \ref{fig: tks_gonogo} shows the size of the clusters using a base-10 log scale for different TDP thresholds. This plot indirectly visualises the topology of the tree $\mathcal{F}$, as it shows the TDP thresholds at which clusters are split into two (or more) subclusters.

\begin{figure}
\centering
\subfloat[Five supra-threshold clusters with CFT $3.2$, and non-zero TDP lower confidence bounds. Brighter colours indicate higher TDP.]{\label{fig_gonogo_1}\includegraphics[width=0.9\textwidth]{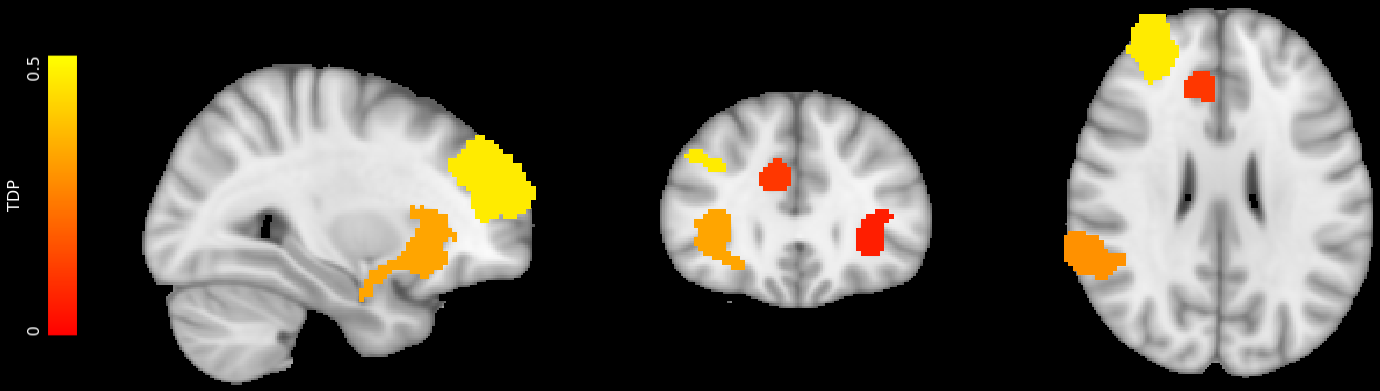}}
\par\medskip
\centering
\subfloat[Five maximal supra-threshold clusters with TDP threshold $\gamma = 0.5$, overlaid on Figure \ref{fig_gonogo_1} in grey. Cluster colours are consistent with Figure \ref{fig: tks_gonogo}.]{\label{fig_gonogo_2}\includegraphics[width=0.9\textwidth]{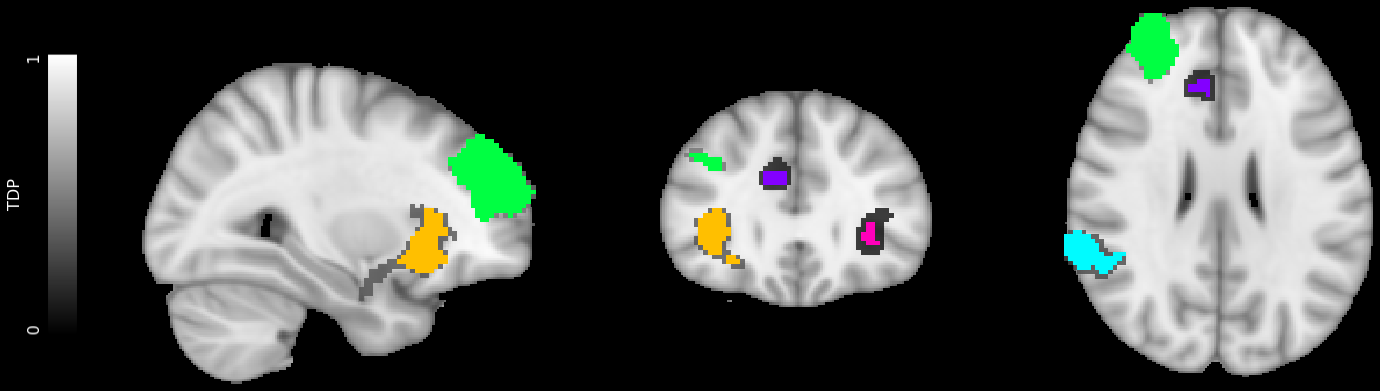}}
\par\medskip
\centering
\subfloat[Four maximal supra-threshold clusters with TDP threshold $\gamma = 0.7$, overlaid on Figure \ref{fig_gonogo_1} in grey. Cluster colours are consistent with Figure \ref{fig: tks_gonogo}.]{\label{fig_gonogo_3}\includegraphics[width=0.9\textwidth]{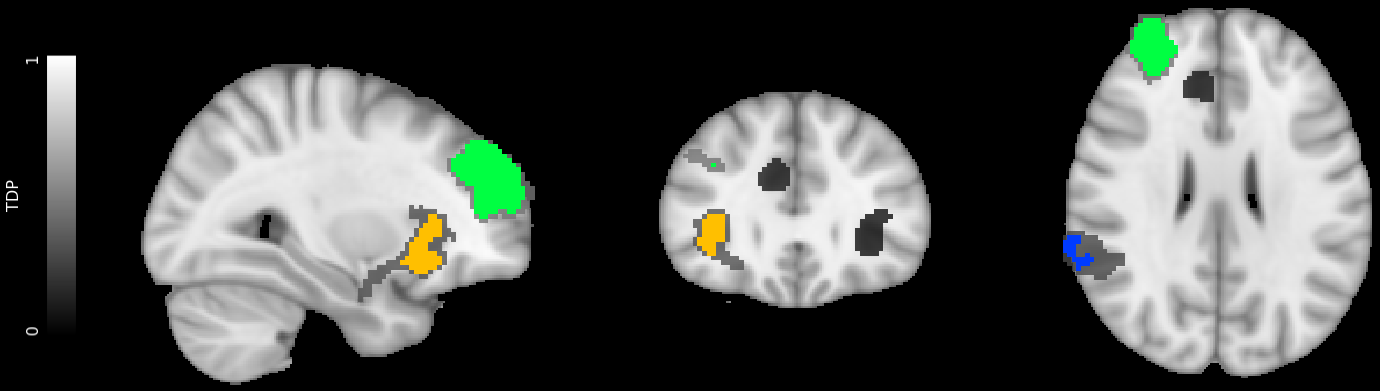}}
\caption{Three families of supra-threshold clusters for the Go/No-go dataset selected according to the criteria specified under each figure.
Each cluster is displayed in a unique colour. The clusters of the last two families are put on top of those of the first family for comparison.}
\label{fig_gonogo}
\end{figure}

We compared the ``adaptive'' supra-threshold clusters we found for $\gamma = 0.5$ and $\gamma = 0.7$ with the conventional supra-threshold clusters that have a constant CFT of $3.2$ and a non-zero TDP lower confidence bound according to ARI. 
See Figure \ref{fig_gonogo} and Table \ref{tbl_gonogo} for details of the comparison. Of the five conventional supra-threshold clusters, all had a TDP confidence bound below 0.5. All clusters found by the adaptive algorithm are subsets of conventional clusters with non-zero TDP lower confidence bounds in the orginal analysis \citep{Rosenblatt2018}. At $\gamma = 0.5$ all these conventional clusters contain a smaller subcluster that fulfils the query. At $\gamma = 0.7$ the two smallest clusters vanish, while the largest cluster breaks into two. In Table \ref{tbl_gonogo}, comparing results for $\gamma = 0.5$ and $\gamma = 0.7$, we note that the cluster in the right frontal pole retains its size quite well as $\gamma$ is increased.

\begin{sidewaystable}[!htbp]
\caption{Cluster attributes of three families of supra-threshold clusters for the Go/No-go dataset. The first family of supra-threshold clusters is obtained in the conventional way by specifying a CFT of 3.2, and a non-zero TDP confidence bound. The other two families of maximal supra-threshold clusters are found by the adaptive thresholding algorithm for TDP thresholds of $0.5$ and $0.7$ respectively.\\
Clusters are categorised by the regions of the brain they belong to. The third family lists two clusters that are part of the first brain region in separate rows. Columns list the size, TDN lower confidence bound, TDP lower confidence bound, and corresponding colour in Figure \ref{fig_gonogo} of each cluster $S$ in a family. 
}
\centering
\begin{threeparttable}
\begin{tabular}{l rrr r rrrr rrrr rrrr}
\toprule
& \multicolumn{3}{c}{MNI coordinates} & \multicolumn{1}{c}{Statistic} & \multicolumn{4}{c}{Conventional} & \multicolumn{4}{c}{Adaptive $(\gamma=0.5)$} & \multicolumn{4}{c}{Adaptive $(\gamma=0.7)$}\\
\cmidrule(lr){2-4} \cmidrule(lr){5-5} \cmidrule(lr){6-9} \cmidrule(lr){10-13} \cmidrule(l){14-17}
Brain region & $x$ & $y$ & $z$ & $Z_\text{max}$ & $|S|$ & $d(S)$ & $q(S)$ & \rainbowbullet & $|S|$ & $d(S)$ & $q(S)$ & \rainbowbullet & $|S|$ & $d(S)$ & $q(S)$ & \rainbowbullet \\
\midrule
Right MTG/STG, SMG, an- & $50$ & $-26$ & $-6$ & $5.25$ & $2191$ & $624$ & $0.285$ & \colouredbullet{white} & $1248$ & $624$ & $0.500$ & \colouredbullet{color3} & $10$ & $7$ & $0.700$ & \colouredbullet{color3}\tnote{*}\\
\quad gular gyrus&&&&&&&&&&&&& $242$ & $170$ & $0.702$ & \colouredbullet{color4}\\
Right frontal pole & $30$ & $46$ & $38$ & $5.85$ & $1835$ & $847$ & $0.462$ & \colouredbullet{white} & $1694$ & $847$ & $0.500$ & \colouredbullet{color2} & $1210$ & $847$ & $0.700$ & \colouredbullet{color2}\\
Right insular cortex, FOC & $32$ & $20$ & $-10$ & $6.01$ & $1400$ & $454$ & $0.324$ & \colouredbullet{white} & $879$ & $449$ & $0.511$ & \colouredbullet{color1} & $641$ & $449$ & $0.700$ & \colouredbullet{color1}\\
Left insular cortex, FOC & $-32$ & $28$ & $0$ & $5.00$ & $421$ & $25$ & $0.059$ & \colouredbullet{white} & $34$ & $17$ & $0.500$ & \colouredbullet{color6} & & & &\\
Right (para)cingular gyrus & $8$ & $22$ & $40$ & $4.92$ & $304$ & $33$ & $0.109$ & \colouredbullet{white} & $66$ & $33$ & $0.500$ & \colouredbullet{color5} & & & &\\
\midrule
Total & & & & & $7513$ & $1983$ & $0.264$ & & $3921$ & $1970$ & $0.502$ & & $2103$ & $1473$ & $0.700$ &\\
\bottomrule\addlinespace[1ex]
\end{tabular}
\begin{tablenotes}\footnotesize
\item[*] This tiny cluster cannot be visualised in Figure \ref{fig_gonogo_3} as it lies in another slice of the brain. Its colour is consistent with Figure~\ref{fig: tks_gonogo}.
\end{tablenotes}
\end{threeparttable}
\label{tbl_gonogo}
\end{sidewaystable}


\subsection{Auditory dataset}

The Auditory dataset contains 218 healthy subjects who were tasked with distinguishing between vocal and non-vocal sounds. It was collected by \citet{Pernet2015}, and is currently freely accessible on OpenNeuro \citep{Poldrack2013} (OpenNeuro Dataset ds000158). Analogous to the Go/No-go dataset, the acquired task-relevant brain images with auditory stimuli were analysed using FSL \citep{Smith2004,Jenkinson2012}. We extracted a subsample of 33 subjects so that the sample size was comparable with the Go/No-go study, and derived the Vocal$>$Non-vocal contrasts with a group-level analysis, consistent with \citet{Rosenblatt2018}. For further information on the experiment and image preprocessing, we refer to \citet{Pernet2015} and \citet{Rosenblatt2018}.

\begin{figure}
\begin{center}
\includegraphics[trim = 0mm 0mm 0mm 0mm, clip=true, width=\textwidth]{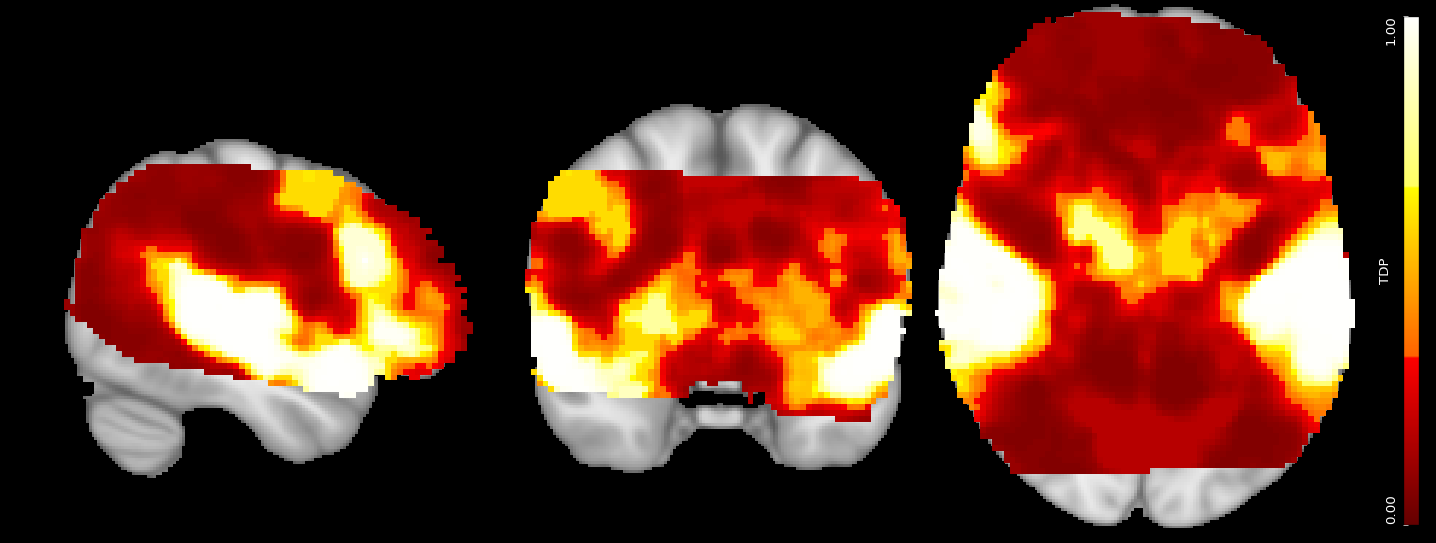}
\caption{A heat map for the Auditory dataset that shows for each voxel $v$ the maximum TDP threshold $\gamma_v$ for which $v$ appears in a supra-threshold cluster $S$ with $q(S) \geq \gamma_v$. Brighter colours correspond to higher TDP thresholds.}
\label{fig: grad_auditory}
\end{center}
\end{figure}

\begin{figure}[ht]
\begin{center}
\includegraphics[trim = 0mm 0mm 0mm 0mm, clip=true, width=\textwidth]{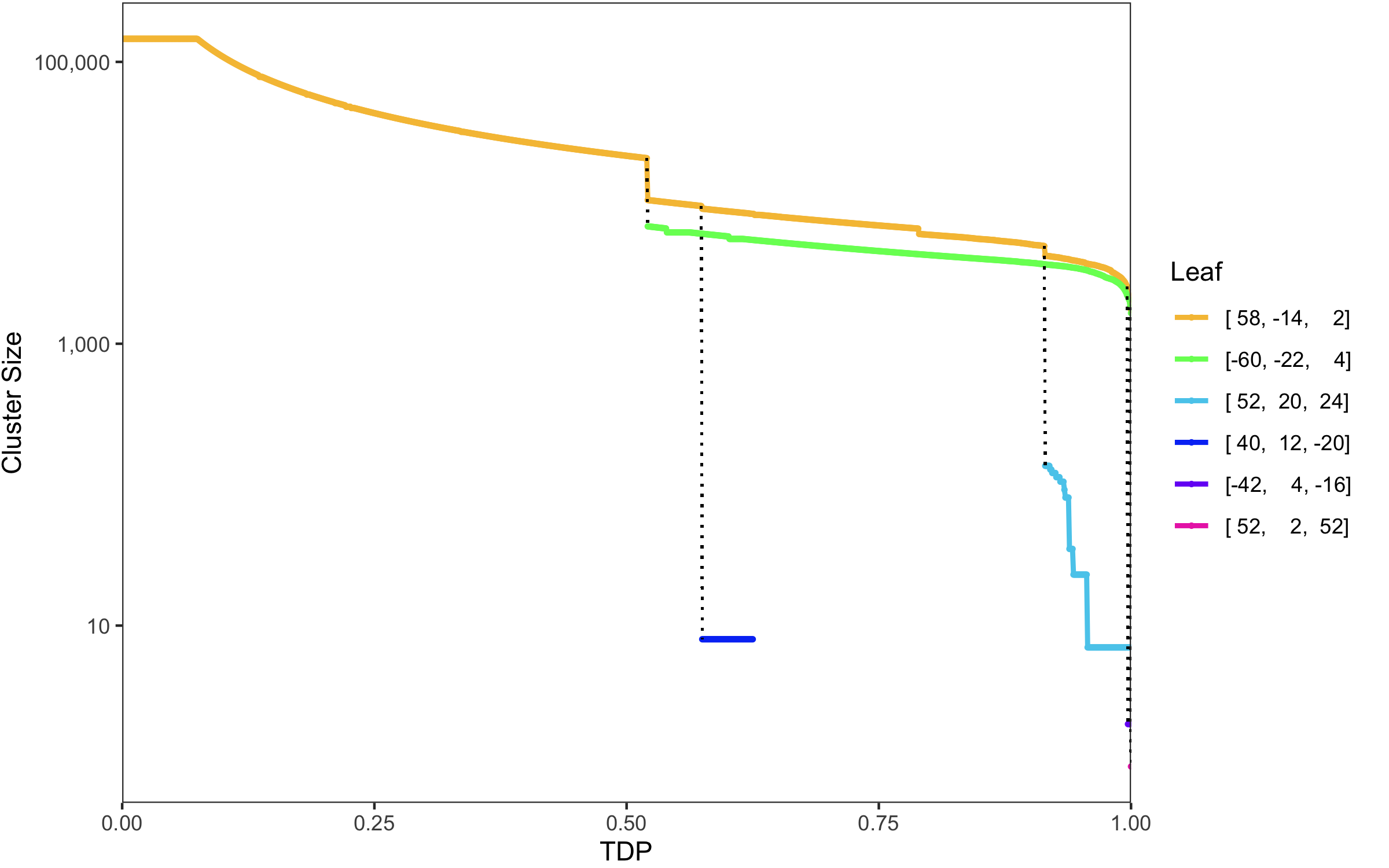}
\caption{A line plot for the Auditory dataset that shows how the size of adaptive maximal supra-threshold clusters (on a log scale) varies with the TDP threshold. Clusters are categorised by the MNI coordinates of the end of the heavy path through the cluster's largest vertex. Dashed lines indicate cluster separation, where a cluster splits into the subclusters connected by the dashed line for more stringent TDP thresholds.}
\label{fig: tks_auditory}
\end{center}
\end{figure}

Constructing the adaptive thresholding data structure took about 0.2 seconds for $m=145872$ in-mask voxels. Using this data structure, it took an average of 0.006 seconds for an average output size of 36640 voxels to handle a query for maximal supra-threshold clusters for each TDP threshold $\gamma \in \{0, 0.01, 0.02, \ldots, 1\}$. See Figure \ref{fig: grad_auditory} for an impression of the maximal supra-threshold clusters at any TDP threshold, and Figure \ref{fig: tks_auditory} for an idea of their cluster size on a logarithmic scale. Compared with Figure \ref{fig: grad_gonogo}, Figure \ref{fig: grad_auditory} tends to give larger cluster sizes for every TDP, which is expected since it is known that auditory stimuli tend to induce stronger activation relative to cognitive tasks.

\begin{figure}
\centering
\subfloat[Three supra-threshold clusters with CFT 3.2, and non-zero TDP lower confidence bounds. Brighter colours indicate higher TDP.]{\label{fig_auditory_1}\includegraphics[width=0.9\textwidth]{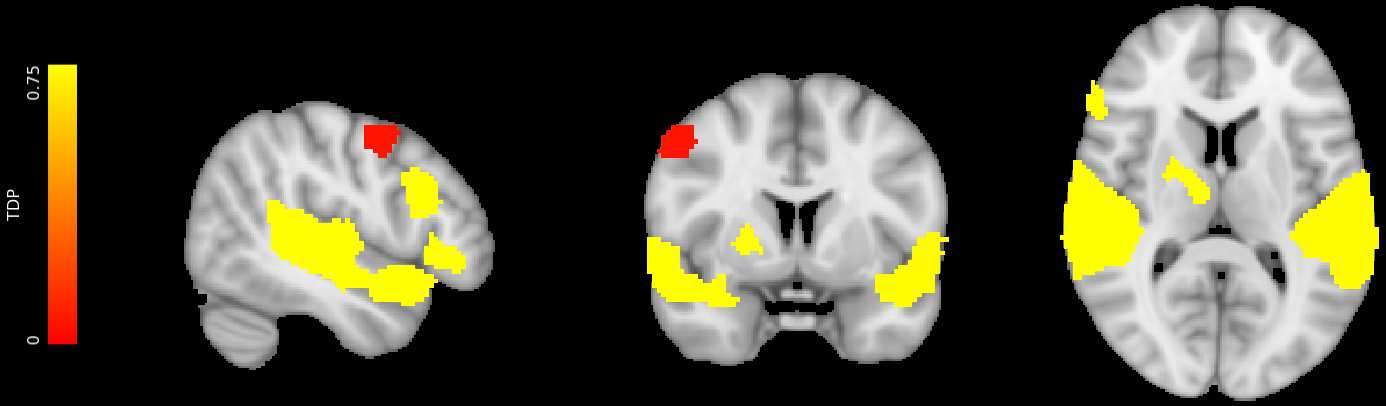}}
\par\medskip
\centering
\subfloat[Two maximal supra-threshold clusters with TDP threshold $\gamma = 0.7$, overlaid on Figure \ref{fig_auditory_1} in grey. Cluster colours are consistent with Figure \ref{fig: tks_auditory}.]{\label{fig_auditory_2}\includegraphics[width=0.9\textwidth]{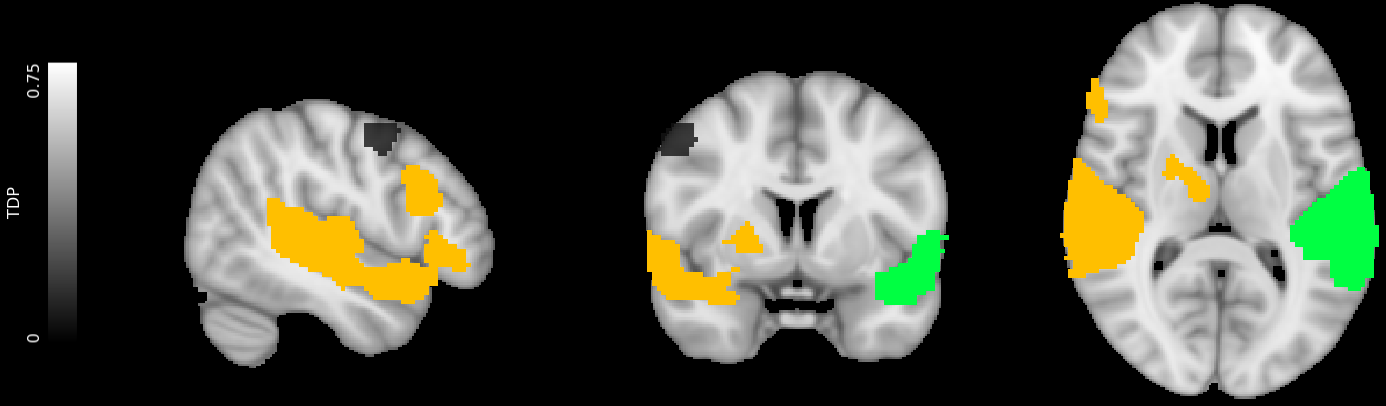}}
\par\medskip
\centering
\subfloat[Two maximal supra-threshold clusters with TDP threshold $\gamma = 0.9$, overlaid on Figure \ref{fig_auditory_1} in grey. Cluster colours are consistent with Figure \ref{fig: tks_auditory}.]{\label{fig_auditory_3}\includegraphics[width=0.9\textwidth]{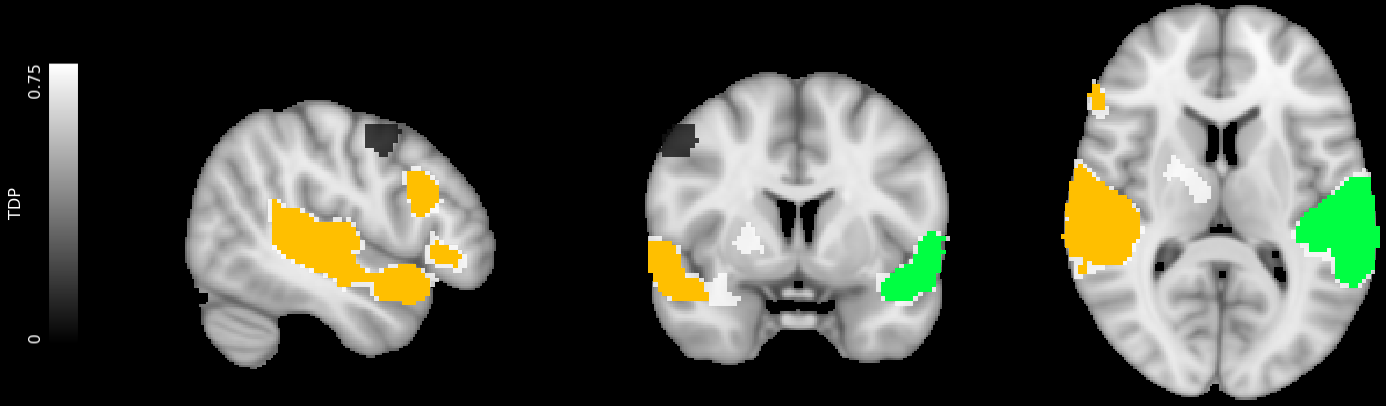}}
\caption{
Three families of supra-threshold clusters for the Auditory dataset selected according to the criteria specified under each figure. Each cluster is displayed in a unique colour. The clusters of the last two families are put on top of those of the first family for comparison.}
\label{fig_auditory}
\end{figure}

As before, we compared the adaptive supra-threshold clusters for TDP thresholds $\gamma = 0.7$ and $\gamma = 0.9$ to the conventional supra-threshold clusters formed by applying a fixed CFT of $Z \ge 3.2$ and filtering out clusters with zero TDP bounds.
Note that we used slightly higher values for $\gamma$ as the amount of signal in the Auditory dataset was much higher than in the Go/No-go dataset. See Figure \ref{fig_auditory} and Table \ref{tbl_auditory} for details about the three families of supra-threshold clusters.

\begin{sidewaystable}[!htbp]
\caption{Cluster attributes of three families of supra-threshold clusters for the Auditory dataset. The first family of supra-threshold clusters is obtained in the conventional way by specifying a CFT of 3.2, and a non-zero TDP confidence bound. The other two families of maximal supra-threshold clusters are found by the adaptive thresholding algorithm for TDP thresholds of $0.7$ and $0.9$ respectively.\\
Clusters are categorised by the regions of the brain they belong to. Columns list the size, TDN lower confidence bound, TDP lower confidence bound, and corresponding colour in Figure \ref{fig_auditory} of each cluster $S$ in a family. 
}
\centering
\begin{threeparttable}
\begin{tabular}{l rrr r rrrr rrrr rrrr}
\toprule
& \multicolumn{3}{c}{MNI coordinates} & \multicolumn{1}{c}{Statistic} & \multicolumn{4}{c}{Conventional} & \multicolumn{4}{c}{Adaptive $(\gamma=0.7)$} & \multicolumn{4}{c}{Adaptive $(\gamma=0.9)$}\\
\cmidrule(lr){2-4} \cmidrule(lr){5-5} \cmidrule(lr){6-9} \cmidrule(lr){10-13} \cmidrule(l){14-17}
Brain region & $x$ & $y$ & $z$ & $Z_\text{max}$ & $|S|$ & $d(S)$ & $q(S)$ & \rainbowbullet & $|S|$ & $d(S)$ & $q(S)$ & \rainbowbullet & $|S|$ & $d(S)$ & $q(S)$ & \rainbowbullet\\
\midrule
Right HG/STG/IFG/PT & $58$ & $-14$ & $2$ & $7.83$ & $6907$ & $5179$ & $0.750$ & \colouredbullet{white} & $7398$\tnote{*} & $5179$ & $0.700$ & \colouredbullet{color1} & $5077$ & $4570$ & $0.900$ & \colouredbullet{color1}\\
Left HG/STG/PT & $-60$ & $-22$ & $4$ & $7.51$ & $4607$ & $3409$ & $0.740$ & \colouredbullet{white} & $4870$\tnote{$\dagger$} & $3409$ & $0.700$ & \colouredbullet{color2} & $3753$ & $3378$ & $0.900$ & \colouredbullet{color2}\\
Right precentral gyrus & $52$ & $2$ & $52$ & $4.88$ & $249$ & $15$ & $0.060$ & \colouredbullet{white} & & & & & & & &\\
\midrule
Total & & & & & $12316$ & $8603$ & $0.699$ & & $12268$ & $8588$ & $0.700$ & & $8830$ & $7948$ & $0.900$ & \\
\bottomrule\addlinespace[1ex]
\end{tabular}
\begin{tablenotes}\footnotesize
\item[*] This cluster consists of $6907$ voxels with $Z \ge 3.2$, and $491$ voxels with $Z < 3.2$.
\item[$\dagger$] This cluster consists of $4607$ voxels with $Z \ge 3.2$, and $263$ voxels with $Z < 3.2$.
\end{tablenotes}
\end{threeparttable}
\label{tbl_auditory}
\end{sidewaystable}


We compared the ``adaptive'' supra-threshold clusters we found for $\gamma = 0.7$ and $\gamma = 0.9$ with the conventional supra-threshold clusters that have a constant CFT of $3.2$ and a non-zero TDP lower confidence bound based on ARI.
See Figure \ref{fig_auditory} and Table \ref{tbl_auditory} for details of the comparison. Of the three  conventional supra-threshold clusters, two have a TDP confidence bound slightly above 0.7 and the other one has a very small TDP confidence bound. As a consequence, the two largest conventional clusters are subsets of the clusters found by the adaptive algorithm for $\gamma = 0.7$. At that $\gamma$, the two big conventional clusters are enlarged by including voxels with $z$-score below 3.2 to form the adaptive clusters, while the smallest conventional cluster with a negligible TDP bound vanishes. At $\gamma = 0.9$ all adaptive supra-threshold clusters are subclusters of the conventional clusters.

\section{Discussion}

In this paper we proposed an efficient, partially online (meaning its input only needs to be partially given at the start), and output-sensitive (meaning its running time scales with the output size) algorithm to find all maximal supra-threshold clusters whose TDP lower confidence bounds meet or exceed a given threshold, for any number of thresholds that need not be specified beforehand. It complements ARI \citep{Rosenblatt2018} by allowing researchers to obtain clusters for a given TDP, rather than TDPs for given clusters. In fact, the algorithm returns lower confidence bounds for all supra-threshold clusters simultaneously, and these clusters and bounds are structured in a forest that can easily be navigated, e.g.,\ using interactive software. Since ARI's lower confidence bound is simultaneous over all clusters in the forest, cherry-picking of clusters with good enough size and high enough TDP does not invalidate error control. The question what value of TDP is high enough for a cluster to be worth reporting should be answered by the field, which relates to the localisation accuracy trade-off and depends on the amount of expected signal. While any non-zero TDP is indicative of the presence of some signal in the cluster, TDP bounds of 40\%, 70\%, and 90\% could be characterised as weak, moderate, and strong spatial localisation, respectively.

We point out that our adaptive thresholding algorithm is not specific to a neuroimaging context, but can be used in general when relations between hypotheses are dictated by a graph, which does not have to be connected. For dense graphs, the task of finding all supra-thresholds clusters dominates the computation time, whereas for sparse graphs (as in neuroimaging), computing the TDP lower confidence bounds of every supra-threshold cluster is the dominant factor.

We also note that the framework we developed is more flexible than querying for maximal supra-threshold clusters with big enough TDP lower confidence bounds. Since a TDP lower confidence bound also yields a lower confidence bound on the number of true discoveries (and vice versa), we are free to pick supra-threshold clusters post-hoc on criteria involving those confidence bounds as well as cluster properties such as size, location, and shape, and do so as many times as desired. Instead of filtering supra-threshold clusters on a TDP threshold, we could take a user-defined predicate to filter clusters by. And rather than selecting the supra-threshold clusters that are maximal with respect to set inclusion from the remaining clusters, we could consider maximality with respect to a user-defined partial order on supra-threshold clusters that induces the same comparability relation as set inclusion. This condition imposed on the partial order ensures that the maximal clusters can simply be computed using a traversal of the directed rooted forest from Section \ref{sec: cluster indexing}. Particularly, in addition to the TDP threshold, we could also account for minimal size threshold and prioritise certain anatomical regions. The output-sensitive quality will generally be lost for the resulting algorithm, however, because we would not know what type of queries to prepare for.

The forest representation of all supra-threshold clusters, calculated by Algorithm \ref{algA}, may also be useful in algorithmic tasks in neuroimaging not related to TDP. They could be used for efficient implementation in methods that involve supra-threshold clusters for several thresholds, e.g.,\ permutation-based cluster-extent thresholding \citep{Nichols2002,Hayasaka2003,Hayasaka2004}, resampling-based clustering \citep{Cox2019}, or threshold-free cluster enhancement \citep{Smith2009}.


Finally, we remark that both ARI and the method presented here are based on the Simes test and can be modified for hypothesis tests that resemble the Simes test. Examples of such tests include the more conservative test by \citet{Hommel1983} that is valid regardless of any dependence between the $p$-values, the tests proposed by \citet{Donoho2004, Blanchard2008}, as well as the tests implied by the permutation-based variants of ARI by \cite{Andreella2020} and \cite{Blain2022}.

\section*{Author contributions}

Alphabetic author order was used. XC, JG and WW conceived the project. TK proposed the use of Algorithm A, improving an earlier proposal by XC and JG. TK and RM designed Algorithm B and proved Theorem 1. TK proved Theorems 2 and 3 and proposed the algorithms in Sections 3.3 and 3.4. XC, TK and RM wrote the software. XC and WW analyzed the data. XC, JG and TK wrote the paper.

\section*{Acknowledgements}

This work was supported by Nederlandse Organisatie voorWetenschappelijk Onderzoek (Grant/Award Number: 639.072.412). We thank Jan Ruitenbeek, who helped to refloat the project when it was stuck.

\clearpage
\appendix
\appendixpage

\algBCorrectness*
\begin{proof}
Algorithm \ref{algB} clearly maintains the invariant that $\mathcal D$ is a partition of $[\ell]$ into integer intervals, on which we impose the usual interval order (i.e., $I < J$ for integer intervals $I$ and $J$ if and only if $\max I < \min J$, and $I \leq J$ if and only if $I < J$ or $I = J$). Let $\mathcal D_0$ be the initial value of $\mathcal D$ in Algorithm \ref{algB}, and let $\mathcal D_i$ be its value after the $i$-th iteration of the for-loop. 
We show by induction on $i$ that 
\[
f_i(k) = d_i - \lvert\{L \in \mathcal D_i : L < K\} \rvert
\]
for all $i \in [\ell]$, $K \in \mathcal D_i$, and $k \in K$, from which it follows that $d_i = f_i(1) = d(V_i)$ for $i \in [\ell]$.
The induction hypothesis certainly holds for $i = 0$, so let us prove it for $i$ while assuming it is true for $i-1$.

If $c(v_i) > \ell$, then $f_i = f_{i-1}$, $d_i = d_{i-1}$, and $\mathcal D_i = \mathcal D_{i-1}$, so the induction hypothesis follows from induction on $i-1$.

Otherwise, let $I \in \mathcal D_{i-1}$ contain $c(v_i)$. Pick any $K \in \mathcal D_{i-1}$ and any $k \in K$. By induction on $i-1$ we have $f_{i-1}(k) = f_{i-1}(\max K)$ and $f_{i-1}(\max K + 1) = f_{i-1}(\max K) - 1$, so
\[
f_{i-1}(k) = \delta(V_{i-1}, \max K) \geq \delta(V_{i-1}, k).
\]
Using Iverson brackets we get
\begingroup
\allowdisplaybreaks
\begin{align}
f_i(k)
&= \max_{j \in \{k, k+1, \ldots, \ell\}} \delta(V_i, j) \notag\\
&= \max_{j \in \{k, k+1, \ldots, \ell\}} \Bigl( \delta(V_{i-1}, j) + [c(v_i) \leq j] \Bigr) \notag\\
&= \max \Bigl\{\delta(V_{i-1}, \max J) + [c(v_i) \leq \max J] \;\Big\vert\; J \in \mathcal D_{i-1},\, J \geq K\Bigr\} \notag\\
&= \max \Bigl\{d_{i-1} - \lvert\{L \in \mathcal D_{i-1} : L < J\} \rvert + [I \leq J] \;\Big\vert\; J \in \mathcal D_{i-1},\, J \geq K\Bigr\} \notag\\
&= d_{i-1} - \lvert\{L \in \mathcal D_{i-1} : L < K\} \rvert + [I \leq K].
\label{eq:delta-recurrence}
\end{align}
\endgroup

If $\min I = 1$, then $[I \leq K] = 1$, $d_i = d_{i-1} + 1$, and $\mathcal D_i = \mathcal D_{i-1}$, so the induction hypothesis follows from Equation \eqref{eq:delta-recurrence}.

Otherwise, let $J \in \mathcal D_{i-1}$ contain $\min I -1$. Continuing from Equation \eqref{eq:delta-recurrence} we get, using $d_i = d_{i-1}$ and $\mathcal D_i = \bigl(\mathcal D_{i-1} \setminus \{I, J\}\bigr) \cup \{I \cup J\}$, that
\begin{align*}
f_i(k)
&= d_i - \lvert\{L \in \mathcal D_i : L < K\} \rvert + [I \cup J < K] - [I < K] - [J < K] + [I \leq K] \\
&= d_i - \lvert\{L \in \mathcal D_i : L < K\} \rvert.
\end{align*}
In particular, if $K \in \{I, J\}$ we also find
\begin{align*}
f_i(k) 
&= d_i - \lvert\{L \in \mathcal D_i : L < K\} \rvert \\
&= d_i - \lvert\{L \in \mathcal D_i : L < I \cup J\} \rvert.
\end{align*}
Hence, the induction hypothesis follows.
\end{proof}

\optimalpathcovers*

\begin{proof}
Let $\mathcal P$ be optimal. 

If $\mathcal P$ is not vertex-disjoint, then there are distinct paths $P,Q \in \mathcal P$ that share a vertex. Let $u$ be the first vertex on $P$ they have in common, which must be the starting vertex of $P$ or $Q$. Without loss of generality, assume $P$ starts at $u$. If $P$ ends at $u$, then we arrive at the contradiction that removing $P$ from $\mathcal P$ yields a path cover $\mathcal P'$ of $\mathcal F$ that satisfies $\sigma(\mathcal F, \mathcal P') = \sigma(\mathcal F) - \lvert\mathcal F_u\rvert < \sigma(\mathcal F)$. Otherwise, $u$ has a child $v$ that lies on $P$, so we arrive at the contradiction that removing $u$ from $P$ yields a path cover $\mathcal P'$ of $\mathcal F$ that satisfies $\sigma(\mathcal F, \mathcal P') = \sigma(\mathcal F) - \lvert\mathcal F_u\rvert + \lvert\mathcal F_v\rvert < \sigma(\mathcal F)$.

If $\mathcal P$ is not minimal, then there is a path $P \in \mathcal P$ that does not end in a leaf of $\mathcal F$, because a minimal path cover of $\mathcal F$ has as many paths as $\mathcal F$ has leaves. Pick a child $v$ of the last vertex on $P$. Since $\mathcal P$ is vertex-disjoint, there is a path $Q \in \mathcal P$ that starts at $v$. But then we can replace $P$ and $Q$ by their join to obtain a path cover $\mathcal P'$ of $\mathcal F$ that satisfies $\sigma(\mathcal F, \mathcal P') = \sigma(\mathcal F) - \lvert\mathcal F_v\rvert < \sigma(\mathcal F)$.

Now pick any minimal, vertex-disjoint, heavy path cover $\mathcal Q$ of $\mathcal F$.
As both $\mathcal P$ and $\mathcal Q$ are minimal and vertex-disjoint, each non-leaf in $\mathcal F$ has exactly one outgoing edge taken by a path in $\mathcal P$ and one taken by a path in $\mathcal Q$. Suppose that some path $P \in \mathcal P$ takes an edge $(u,v)$, but some path in $\mathcal Q$ takes a different edge $(u,w)$. Let $P^u$ be the subpath of $P$ from start to $u$, let $P_v$ be the subpath of $P$ from $v$ to end, and let $P_w \in \mathcal P$ be the path that starts at $w$. By replacing $P$ and $P_w$ in $\mathcal P$ by $P_v$ and the join of $P^u$ and $P_w$, we construct another minimal, vertex-disjoint path cover $\mathcal P'$. Since $(u,w)$ is heavy, we have
\[\sigma(\mathcal F) \leq \sigma(\mathcal F, \mathcal P') = \sigma(\mathcal F) + \lvert\mathcal F_v\rvert - \lvert\mathcal F_w\rvert \leq \sigma(\mathcal F).\]
It follows that $\lvert\mathcal F_v\rvert = \lvert\mathcal F_w\rvert$, so $(u,v)$ is heavy, and therefore $\mathcal P$ is too. It also follows that $\mathcal P'$ is optimal, so we may repeat this exchange argument to transform $\mathcal P'$ into $\mathcal Q$ while preserving optimality.
\end{proof}

\worstcasesigma*

\begin{proof}
We first show the upper bound $\sigma(m) \leq f(m)$ by induction on $m$, where $f\colon \mathbb [0,\infty) \to \mathbb R$ is the convex function given by $f(x) = x \log_4 x + x$ for $x > 0$ and $f(0) = \lim_{x \rightarrow 0^+} f(x) = 0$.

Let $\mathcal F$ be a directed rooted forest of order $m$ such that $\sigma(m) = \sigma(\mathcal F)$, and let $\mathcal P$ be an optimal path cover of $\mathcal F$. We may assume that $\mathcal F$ is a tree without loss of generality. Namely, if $u$ and $v$ are distinct roots of $\mathcal F$ at which paths $P_u, P_v \in \mathcal P$ start, we may create a new directed rooted forest $\mathcal F'$ from $\mathcal F$ by adding an edge from the end vertex of $P_u$ to $v$, and give it the optimal path cover $\mathcal P'$ created from $\mathcal P$ by joining $P_u$ to $P_v$. Then $\mathcal F'$ has one less root than $\mathcal F$ and $\sigma(\mathcal F') = \sigma(\mathcal F) - \lvert\mathcal F_u\rvert - \lvert\mathcal F_v\rvert + \lvert\mathcal F_u'\rvert =  \sigma(\mathcal F)$.

If $m = 1$, then $\sigma(1) = 1 = f(1)$. Now assume $m \geq 2$ and that the induction hypothesis $\sigma(m') \leq f(m')$ holds for all positive integers $m' < m$. Let $u$ be the root of $\mathcal F$, let $N(u)$ be the set of children of $u$, and let $(u,v)$ be a heavy edge taken by a path in $\mathcal P$. Then we find by induction that
\begin{align}
\sigma(m) 
&= \sigma\bigl(\mathcal F[V \setminus \{u\}]\bigr) + \lvert\mathcal F_u\rvert - \lvert\mathcal F_v\rvert \notag\\
&= \sum_{w \in N(u)} \sigma(\mathcal F_w) + m - \lvert\mathcal F_v\rvert \notag\\
&\leq \sum_{w \in N(u)} f\bigl(\lvert\mathcal F_w\rvert\bigr) + m - \lvert\mathcal F_v\rvert. \label{eq:sigmam}
\end{align}
Since $(u,v)$ is heavy, the sequence $\bigl(\lvert\mathcal F_w\rvert\bigr)_{w \in N(u)}$ is majorised by the sequence
\[\Bigl(\underbrace{\lvert \mathcal F_v\rvert,\, \lvert \mathcal F_v\rvert,\,\ldots,\, \lvert \mathcal F_v\rvert}_{k\text{ times}},\, m-1-k\lvert \mathcal F_v\rvert\Bigr) \]
for $k = \bigl\lfloor\frac{m-1}{\lvert\mathcal F_v\rvert}\bigr\rfloor \geq 1$ after making both sequences equally long by padding with zeros if necessary. 
Recall here that a sequence $(a_i)_{i=1}^n$ \emph{majorises} a sequence $(b_i)_{i=1}^n$ if $\sum_{i=1}^j \vec{a}_i \geq \sum_{i=1}^j \vec{b}_i$ for all $j \in [n]$ and $\sum_{i=1}^n a_i = \sum_{i=1}^n b_i$, where $\vec{a}$ (resp.\ $\vec{b}$) lists the elements of $a$ (resp.\ $b$) in descending order.
In particular, we have $\frac{m-1}{k+1} < \lvert \mathcal F_v \rvert \le \frac{m-1}{k}$.

The aforementioned majorisation lets us apply Karamata's inequality to Equation \eqref{eq:sigmam} to get that
\[
\sigma(m) \leq g_k\bigl(\lvert\mathcal F_v\rvert\bigr),
\]
where $g_j\colon \bigl[\frac{m-1}{j+1}, \frac{m-1}{j}\bigr] \to \mathbb R$ is the function given by \[g_j(x) = jf(x) + f(m-1-jx) + m-x\] for $j \in \{k, k+1\}$. We also observe that $g_k(x) = g_{k+1}(x)$ for $x = \frac{m-1}{k+1}$. Since $g_k$ is a sum of convex functions, it is convex itself, so it attains its maximum on the boundary of its domain. Using additionally that $\log_4(x) \leq x-1$ for $x \in \bigl(0,\infty\bigr) \setminus \bigl(\frac{1}{2},1\bigr)$, we conclude that
\begingroup
\allowdisplaybreaks
\begin{align*}
\sigma(m) 
&\leq \max\bigl\{g_k\bigl(\tfrac{m-1}{k+1}\bigr),\, g_k\bigl(\tfrac{m-1}{k}\bigr)\bigr\} \\
&= \max_{j\in\{k,k+1\}} g_j\bigl(\tfrac{m-1}{j}\bigr) \\
&= \max_{j\in\{k,k+1\}} \Bigl(f(m-1) + 1 + (m-1)\bigl(1-\tfrac{1}{j} + \log_4 \tfrac{1}{j}\bigr)\Bigr) \\
&\leq f(m-1) + 1 \\
&\leq f(m),
\end{align*}
\endgroup
establishing the induction hypothesis.

We show the lower bound of $\sigma(m)$ by considering $\sigma(\mathcal F)$ for a complete binary tree $\mathcal F$ of order $m$. Recall that vertices in a binary tree have at most two children, called a \emph{left} and a \emph{right} child. We may uniquely label vertices of a binary tree by
labelling its root with $1$, and labelling the left and right children of a vertex that has label $i$ with labels $2i$ and $2i+1$ respectively (provided they exist). Recall then that a binary tree of order $m$ is \emph{complete} if its vertices are labelled by $[m]$. See for example below for a complete binary tree of order $6$.
\begin{center}
\begin{tikzpicture}[every node/.style={circle,draw, inner sep=0pt, minimum size=15pt}, scale=1, >={Stealth[round]}]
	\node(1) at (0,3) {$1$}; 
	\node(2) at (-1,2) {$2$}; 
	\node(3) at (1,2) {$3$};
	\node(4) at (-1.5,1) {$4$}; 
	\node(5) at (-.5,1) {$5$}; 
	\node(6) at (.5,1) {$6$};
	
	\draw[->] (1) edge (2) edge (3);
	\draw[->] (2) edge (4) edge (5);
	\draw[->] (3) edge (6);
\end{tikzpicture}
\end{center}

Observe that $\mathcal F$ has an optimal path cover $\mathcal P$ whose paths start at the odd-labelled vertices and continue to even-labelled vertices. The example tree above has an optimal path cover consisting of the three paths from 1 to 4, from 3 to 6, and from 5 to 5. Therefore, 
\begingroup
\allowdisplaybreaks
\begin{align*}
\sigma(\mathcal F, \mathcal P) 
&= \sum_{v \in I(\mathcal P)} \lvert\mathcal F_v \rvert \\
& = \sum_{\substack{0 \leq i \leq m\\ i \equiv 1\mkern-14mu \pmod 2}} \sum_{\substack{k \geq 0,\, 0 \leq j < 2^k\\i2^k + j \leq m}} 1 \\
& = \sum_{k\geq 0} \sum_{\substack{0 \leq i \leq m\\ 0 \leq j < 2^k\\i2^k + j \leq m}} [i \equiv 1\mkern-14mu \pmod 2] \\
& = \sum_{k\geq 0} \sum_{0 \leq n\leq m} \bigl[\lfloor n/2^k\rfloor \equiv 1\mkern-14mu \pmod 2\bigr] \tag{Set $n = i2^k+j$} \\
&= \sum_{0 \leq n\leq m} s_2(n) \\
&= m \log_4 m + O(m),
\end{align*}
\endgroup
where $s_2(n) = \sum_{k=0}^{\infty} a_k(n)$ is the sum of the bits of $n$ when uniquely expressed in binary as $n = \sum_{k=0}^{\infty} a_k(n) 2^k$ with all $a_k(n) \in \{0,1\}$, and where the last step is a result due to \citep{Bush1940,Bellman1948,Mirsky1949}. For more on the sequence $S_2(m) = \sum_{n=0}^m s_2(n)$, see \cite[seq.\ A000788]{OEISA000788}.

\begin{remark} 
Numerical calculations up to $m \leq 10^4$ suggest that \[\sigma(m) \leq (m+1) \log_4(m+1) +  (\tfrac{5}{6} - \log_{4} 3)(m+1)\] for any $m$, where the linear term has a small constant $\frac{5}{6} - \log_{4} 3 \approx 0.041$. It is not hard to verify that this bound is tight for the trees $\mathcal F'$ of order $m = 3\cdot 2^k - 1$ that arise by attaching four new children to each leaf of a complete binary tree of order $2^k - 1$. \qedhere
\end{remark}
\end{proof}



\clearpage
\bibliography{SPMalgorithm}
\bibliographystyle{apalike}

\end{document}